\documentclass[12pt]{article}
\usepackage[margin=1.0in]{geometry}
\usepackage[utf8]{inputenc}
\usepackage[T1]{fontenc}
\usepackage{mathptmx}
\usepackage{amsmath,amssymb,amsfonts,mathtools}
\usepackage{bm}
\usepackage{graphicx}
\usepackage{mathrsfs}
\usepackage{bbm}
\usepackage[libertine]{newtxmath}
\pdfmapfile{+newtx.map}
\usepackage{array,booktabs,multirow,float}
\usepackage{xfrac}
\usepackage{slashed}
\usepackage{scalerel}
\usepackage{authblk}
\usepackage[hidelinks]{hyperref}
\usepackage{bookmark}
\usepackage[noabbrev]{cleveref}

\DeclareMathAlphabet{\mathcal}{OMS}{cmsy}{m}{n}

\allowdisplaybreaks

\newcommand{\argu}[1]{(#1)}

\NewDocumentCommand{\ket}{m}{\left|#1\right\rangle}
\NewDocumentCommand{\bra}{m}{\left\langle#1\right|}
\NewDocumentCommand{\matrixel}{m m m}{\langle#1|#2|#3\rangle}
\NewDocumentCommand{\mel}{s s m m m}{\langle#3|#4|#5\rangle}
\NewDocumentCommand{\dblmatel}{m m m}{\left\langle #1\left\Vert #2\right\Vert #3\right\rangle}
\NewDocumentCommand{\dd}{o m}{\mathop{}\!\mathrm{d}\IfValueT{#1}{^{#1}}#2}

\def\NDBD{0\nu\beta\beta}

\def\Smat{\mathcal{S}\text{-matrix}}
\def\Gr{\gamma^{\rho}}
\def\Gs{\gamma^{\sigma}}
\def\Gm{\gamma^{\mu}}

\def\pl{(1-\gamma_5)}
\def\pr{(1+\gamma_5)}
\def\plr{(1\mp\gamma_5)}
\def\prl{(1\pm\gamma_5)}
\def\nn{\nonumber}

\def\intinf{\displaystyle\int}
\def\intinfdd{\intinf\!\!\dd[4]{x_1}\dd[4]{x_2}}
\def\intinfdds{\intinf\!\!\dd[3]{\bm{x}_1}\dd[3]{\bm{x}_2}}
\def\momintomega{\intinf\!\! \frac{\dd[3]{\bm{q}}}{(2\pi)^3}\frac{e^{i\bm{q}\cdot(\bm{x}_1-\bm{x}_2)}}{\omega}}

\def\qqint{\intinf\!\!\frac{\dd[4]{q}}{(2\pi)^4}}
\def\effmass{\langle m_{\beta\beta}\rangle}

\def\({\Big{(}}
\def\){\Big{)}}
\def\[{\Big{[}}
\def\]{\Big{]}}

\begin{document}

\title{Long-range scalar interactions in the effective field theory approach to $\NDBD$}
\author{Fahim Ahmed%
\thanks{Email: \href{mailto:ahmed1f@cmich.edu}
{\nolinkurl{ahmed1f@cmich.edu}}}}
\affil{Department of Physics, Central Michigan University, Mount Pleasant, Michigan 48858, USA}
\maketitle
\begin{abstract}
As a lepton-number-violating process whose observation would establish the
Majorana nature of neutrinos, neutrinoless double-beta decay ($\NDBD$) is an
important low-energy nuclear probe of beyond-the-Standard-Model physics.
Effective field theory (EFT) provides a general, model-independent description, and
interference among effective mechanisms is essential for disentangling their
contributions to the total decay rate. We focus on scalar-type terms of the
long-range Lagrangian in the EFT approach to $\NDBD$. We simultaneously retain the two scalar interactions involving a common $(S+P)$ leptonic current coupled either to an $(S-P)$ or an $(S+P)$ hadronic current. Their coherent contribution produces mutual interference between the two scalar mechanisms, controlled by the relative complex phase of their couplings, in addition to their separate interference with the standard mass mechanism. We give a compact
half-life expression, the explicit scalar--scalar interference coefficient in
terms of scalar nuclear-matrix-element ratios and integrated phase-space
factors, and a positivity bound on that coefficient. The calculation uses the
closure and single-nucleon impulse approximations, retains the dominant scalar
recoil subset, and includes the $s_{1/2}$ and $p_{1/2}$ Coulomb waves of the
outgoing electrons. The resulting dimensionally explicit formulation provides
a basis for correlated analyses of the two scalar coefficients.
\end{abstract}

\section{Introduction}\label{Intro}
Neutrinoless double-beta decay ($\NDBD$) is an important low-energy nuclear process that probes high-energy beyond-the-Standard-Model (BSM) physics. Observation of $\NDBD$ would establish the Majorana nature of neutrinos through the black-box theorem \cite{PhysRevD.25.2951}. Several competing mechanisms have been proposed, many tied to specific BSM models. An effective field theory (EFT) description is more general: its low-energy coefficients can be constrained jointly by $\NDBD$, collider, and other data.

In the EFT approach to $\NDBD$ we consider the long-range part, which is point-like at the Fermi scale and involves the
exchange of a light intermediate neutrino. The most general Lorentz-invariant current$\times$current Hamiltonian of
dimension six that gives rise to the long-range part of $\NDBD$ is \cite{PhysRevC.98.035502,Deppisch2012}
\begin{equation} \label{Ham-longrange}
 \mathcal{H}^{\beta}_6=\frac{G_\beta}{\sqrt{2}}\sum_j\left[U_{ej}j^{\rho j}_{V-A}J^\dagger_{V-A,\rho} + 
\sum^{*}_{A,B}\varepsilon_{A,j}^B j^j_BJ^\dagger_A \right] ,
\end{equation}
where $J^\dagger_A=\overline{u}\mathcal{O}_A d$ and $j^j_B=\overline{e}\mathcal{O}_B \nu_j$ are the hadronic 
and leptonic Lorentz currents, respectively. The effective hadronic currents are between a down-quark ($d$) and an 
up-quark ($u$). Similarly, the leptonic currents signify the interaction between the electron ($e$) and light massive
neutrinos ($\nu_j$). An overbar denotes the Dirac adjoint, for example
$\overline e=e^\dagger\gamma^0$. The $j$ index enumerates the mass eigenstates,
$U_{ej}$ is the electron-row leptonic mixing element, $A$ and $B$ label the
hadronic and leptonic Lorentz structures, and $\varepsilon_{A,j}^{B}$ is a
dimensionless LNV coefficient at the Hamiltonian level. The indices
$\rho$ and $\sigma$ are Lorentz indices and are summed when repeated.
As we are working
in the EFT framework, we do not have to make explicit model-dependent
assumptions about the number of these mass eigenstates; see
Eqs.~\eqref{redefine_eff_param} and \eqref{m_eff}. The effective operators
are combinations of $\gamma$ matrices corresponding to the
Lorentz structures of the currents \cite{Deppisch2012},
\begin{align}
&\text{\textbf{V}ector}\pm\text{\textbf{A}xial-vector type:}\qquad&&\mathcal{O}_{V\pm 
A}=\Gr\left(1\pm\gamma_{5}\right),\label{OV-A}\\ 
&\text{\textbf{S}calar}\pm\text{\textbf{P}seudo-scalar type:}\qquad &&\mathcal{O}_{S\pm P}=\left(1\pm\gamma_{5}\right), \label{OS-P} 
\\
&\text{\textbf{R}ight-\textbf{T}ensor}\pm\text{\textbf{L}eft-\textbf{T}ensor type:}\qquad 
&&\mathcal{O}_{T_{R}/T_{L}}=\sigma^{\rho\sigma}\left(1\pm\gamma_{5}\right)=\tfrac{i}{2}[\Gr,\Gs]\prl .\label{OTRL} 
\end{align}
Here $\gamma_5\equiv i\gamma^0\gamma^1\gamma^2\gamma^3$ and
$\sigma^{\rho\sigma}\equiv\tfrac{i}{2}[\gamma^\rho,\gamma^\sigma]$.
Taking into account all possible combinations of Lorentz structures we have the following set for the lepton number 
violating (LNV) BSM parameters, 
\begin{equation}
\varepsilon_{A,j}^B= \{\varepsilon^{V+A}_{V-A,j}, \varepsilon^{V-A}_{V+A,j}, \varepsilon^{V+A}_{V+A,j}, 
\varepsilon^{S-P}_{S-P,j}, \varepsilon^{S+P}_{S-P,j}, \varepsilon^{S-P}_{S+P,j}, \varepsilon^{S+P}_{S+P,j}, 
\varepsilon^{T_R}_{T_R,j}, 
\varepsilon^{T_R}_{T_L,j}, \varepsilon^{T_L}_{T_R,j}, \varepsilon^{T_L}_{T_L,j}\}. 
\end{equation}
The `*' symbol indicates that the term with $A=B=(V-A)$ is explicitly taken out of the sum\footnote{The context distinguishes the summation label $A$ from the axial-vector label in $V\pm A$.}; this is the term containing two Standard Model (SM) currents. Here $G_\beta=G_F\cos\theta_C$ is the effective point-like coupling between left-handed
quark and left-handed lepton currents, $G_F=1.1663787\times 10^{-5}$ GeV$^{-2}$ is the Fermi constant, and $\theta_C$ is the Cabibbo angle.

The neutrino fields in the leptonic currents $j^j_B=\overline{e}\mathcal{O}_B \nu_j$ are the SM fields, but taken to be massive. We express the massive neutrino fields in the Majorana basis as \cite{Akhmedov:2014kxa},
\begin{align}\label{nu-mass}
 \nu_j=\nu_{jL}+(\nu_{jL})^C,
\end{align}
where $L$ denotes the left-chiral component and the superscript $C$ denotes
charge conjugation. The electron-neutrino flavor field required at the weak
vertices is related to the mass eigenstates by
\begin{align}\label{nu-mixing}
 \nu_{eL}=\sum_{j}U_{ej}\nu_{jL},
\end{align}
where $U_{ej}$ is an element of the Pontecorvo--Maki--Nakagawa--Sakata
(PMNS) matrix. Thus, comparing Eqs.~\eqref{nu-mass} and
\eqref{nu-mixing}, electron neutrinos are linear combinations of the
left-chiral parts of the mass eigenstates. In the well-motivated BSM
scenario of the left-right symmetric model (LRSM), the right-chiral parts
of the mass states ($(\nu_{jL})^C$) are assumed to follow a similar mixing
scheme \cite{Ahmed2020},
\begin{align}
 \nu_{eR}=\sum_j T^*_{ej}(\nu_{jL})^C,
\end{align}
where $T^*_{ej}$ is a right-handed mixing-matrix element and $\nu_{eR}$ is
the right-chiral electron-neutrino flavor field, absent in the SM. Since we
are not focusing on specific models, we use generic parameters for any BSM
couplings. For example, in the LRSM,
$\varepsilon_{V+A, j}^{V+A}$ would be identified with the model parameters as
\begin{align}
 \varepsilon_{V+A, j}^{V+A}\coloneqq  T^*_{ej}\(\frac{g_R}{g_L}\)^2\frac{\[\(\frac{m_{W_1}}{m_{W_2}}\)^2+\tan^2 \xi\]}{\[1+\tan^2 \xi\(\frac{m_{W_1}}{m_{W_2}}\)^2\]},
\end{align}
where $\xi$ is the mixing angle between the gauge eigenstates
$W_L^\pm,W_R^\pm$ and the mass eigenstates $W_1^\pm,W_2^\pm$;
$m_{W_1},m_{W_2}$ are the corresponding masses and $g_L,g_R$ the
left- and right-handed electroweak couplings. See
Refs.~\cite{Ahmed2020,Hirsch:1995rf} for details. Other effective BSM
parameters in Eq.~\eqref{Ham-longrange} can be matched similarly to
specific BSM scenarios \cite{PhysRevD.76.093009}.

In this paper we discuss the interference effects involving the standard mass mechanism and the scalar-pseudoscalar type ($S\pm P$) terms, subsequently referred to as scalar terms for brevity. For general discussions of $\NDBD$ in the EFT approach, see Refs.~\cite{Cirigliano:2018yza,delAguila:2012nu,Deppisch2012}. The full decay-rate expressions for the mass mechanism and the $V\pm A$ terms, including their interference, were derived in Ref.~\cite{Doi+Kotani1985}, and an updated numerical analysis of the corresponding phase-space factors (PSF) and nuclear matrix elements (NME) for several experimentally relevant nuclides was performed in Ref.~\cite{Ahmed2020}. Contributions of the $S\pm P$ and $T_R$ terms to the total decay rate and constraints on the associated effective LNV couplings were studied in Refs.~\cite{PhysRevC.98.035502,PhysRevD.76.093009,PhysRevD.105.099902}. We independently derive the scalar amplitudes and half-life coefficients from the long-range dimension-six Hamiltonian, providing a cross-check of the corrected scalar $P$-wave results in Refs.~\cite{PhysRevD.76.093009,PhysRevD.105.099902}. As extensions, we retain the two scalar amplitudes simultaneously and coherently, derive their mutual scalar--scalar interference and relative-phase dependence in addition to their separate interference with the standard mass mechanism, and express the new interference coefficient in terms of scalar NME ratios and integrated PSF. We also show that the two single-coupling limits recover the corrected results and obtain a positivity bound on the scalar--scalar interference coefficient, thereby providing a consistent basis for correlated analyses of the two scalar interactions.

\section{Formalism for \texorpdfstring{$\NDBD$}{neutrinoless double-beta decay} in the Effective Field Theory approach}\label{chap4-SMat}
We use natural units, $\hslash=c=1$, and the metric
$g^{\mu\nu}=\operatorname{diag}(1,-1,-1,-1)$. Greek indices such as
$\mu,\nu,\rho,$ and $\sigma$ denote spacetime components, while $k$ and
$l$ denote spatial Cartesian components when they occur as tensor indices;
repeated Lorentz and Cartesian indices are summed unless stated otherwise.
Other Latin labels are defined locally at their point of use.
Square-bracket statements of the form $[X]=M^n$ specify the mass dimension
of $X$.
\begin{figure}[htp]
\centering
\includegraphics[width=\textwidth]{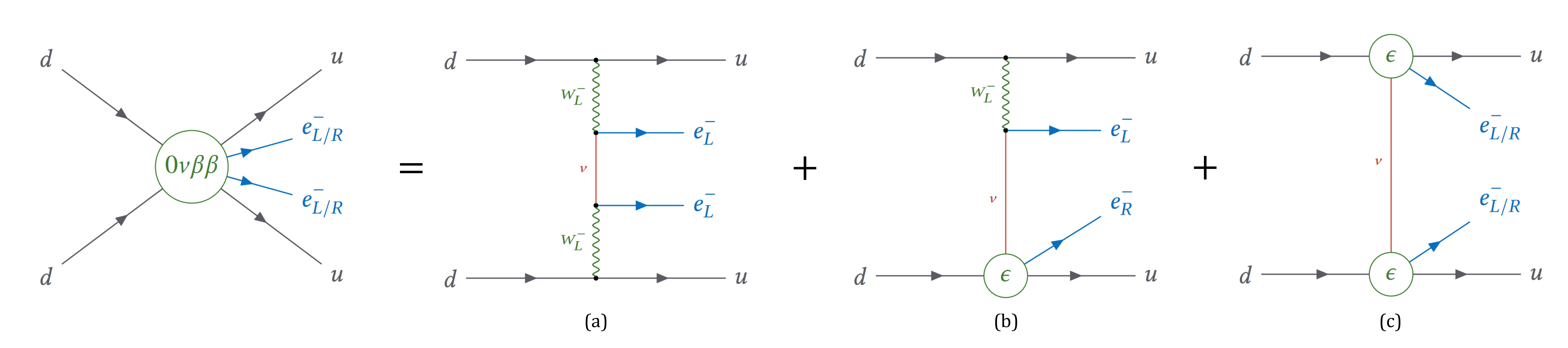}
\caption{Feynman-diagram decomposition of the $\NDBD$ amplitude generated by Eq.~\eqref{Ham-longrange}. The full amplitude (left) equals the sum of (a) the standard mass mechanism with two $V-A$ vertices, (b) a first-order term with one BSM vertex, and (c) a second-order term with two BSM vertices. The first-order coupling $\varepsilon_{A,j}^{B}$ is denoted by $\epsilon$ at the amplitude level; see Eq.~\eqref{redefine_eff_param}. \label{EFT_NDBD}}
\end{figure}
Following the prescription of canonical quantization we consider the second-order term in $G_\beta$ in the S-matrix element for double beta decay ($\beta\beta$) processes,
\begin{align}\label{2ndsx1}
\mathcal S^{(2)}=\frac{(-i)^2}{2!}\int\dd[4]{x_1}\dd[4]{x_2}\,
\mathscr T\!\left[\mathcal H_6^\beta(x_1)\mathcal H_6^\beta(x_2)
\exp\!\left(-i\int\dd[4]{x}\,\mathcal H_S(x)\right)\right].
\end{align}
The first term corresponds to the standard mass mechanism due
to light-neutrino exchange between two SM-like $V-A$ vertices. The third term can be neglected since it
is second order in the small BSM parameters $\varepsilon_{A,j}^B$, see Fig.~\ref{EFT_NDBD}. We thus focus on the phenomenologically interesting first-order terms where one of the vertices is the SM $V-A$ type, as indicated by the $W^-_L$ exchange in Fig.~\ref{EFT_NDBD}(b). The other vertex is of BSM type, and its coefficient is redefined at the amplitude level as $\epsilon$ instead of $\varepsilon$; see Eq.~\eqref{redefine_eff_param}. In Eq.~\eqref{2ndsx1}, $\mathcal H_S$ denotes the strong-interaction Hamiltonian, whose evolution is retained because the nuclear states are strong-interaction eigenstates \cite{Bilenky+Petcov1987}.
The scalar--scalar term derived later is the interference of two such
first-order amplitudes when the rate is formed; it is not the amplitude with
two BSM vertices shown in Fig.~\ref{EFT_NDBD}(c).

Two general cases arise for this first-order term in $\varepsilon_{A,j}^B$
with respect to the product of leptonic currents. As shown in
Sec.~\ref{LME}, the chiral projectors select either a neutrino-mass or a
neutrino-momentum numerator:
\begin{enumerate}
 \item \textbf{Mass-dependent terms:} The leptonic SM $V-A$ current meets a left-handed BSM current $j_B$ with $B=V-A,S-P,T_L$. These contributions are proportional to a light-neutrino mass as well as to an unknown BSM coefficient. We therefore retain only the standard mass mechanism among the mass-dependent terms.
\item \textbf{Momentum-dependent terms:} In the second scenario, the leptonic SM $V-A$ current meets right-handed BSM currents 
 $j_B$ with $B=V+A, S+P, T_R$. These contributions are proportional to the virtual-neutrino momentum $|\bm q|\sim100$ MeV. This
leads to strong constraints on the corresponding effective BSM couplings.
\end{enumerate}
Furthermore, the term for $\varepsilon_{T_L,j}^{T_R}$ can be shown to be identically zero using the identity of the gamma matrices. We focus on the mass-mechanism and scalar terms. Considering the initial nuclear state $\ket{i}=\ket{N_i}$ and final nuclear and lepton state $\ket{f}=\ket{N_f}\ket{e_1^-e_2^-}$, the S-matrix element for $\NDBD$ can be written as
\begin{align}\label{0nmel}
\matrixel{f}{\mathcal{S}_{0\nu}^{(2)}}{i}&=\frac{(-i)^2}{2!}\frac{G^2_{\beta}}{2}\sum_jU_{ej}\intinfdd\nn\\
&\times\Bigg[U_{ej}\matrixel{e_1^-,e_2^-}{\mathscr{T}\left[j^{\rho j}_{V-A}\argu{x_1}j^{\sigma j}_{V-A}
\argu{x_2}\right]}{0} 
\matrixel{N_f}{\mathscr{T}\left[J_{V-A,\rho}^{\dagger}\argu{x_1}J_{V-A,\sigma}^{\dagger}\argu{x_2}\right]}{N_i}\nn\\
&\quad 
+2\varepsilon_{S-P, j}^{S+P}\matrixel{e_1^-,e_2^-}{\mathscr{T}\left[j^{\rho j}_{V-A}\argu{x_1}j^j_{S+P}\argu{x_2}\right]}{0}
\matrixel{N_f}{\mathscr{T}\left[J_{V-A,\rho}^{\dagger}\argu{x_1}J_{S-P}^{\dagger}\argu{x_2}\right]}{N_i}\nn\\
&\quad 
+2\varepsilon_{S+P, j}^{S+P}\matrixel{e_1^-,e_2^-}{\mathscr{T}\left[j^{\rho j}_{V-A}\argu{x_1}j^j_{S+P}\argu{x_2}\right]}{0}
\matrixel{N_f}{\mathscr{T}\left[J_{V-A,\rho}^{\dagger}\argu{x_1}J_{S+P}^{\dagger}\argu{x_2}\right]}{N_i}\Bigg].
\end{align}
The extra factor of 2 in the last two terms inside the square brackets is due to the exchange of vertices.
Here $x_1$ and $x_2$ are the spacetime positions of the two semileptonic vertices, $\mathscr{T}$ denotes time ordering, $\ket{0}$ is the leptonic vacuum, and $\mathcal{S}_{0\nu}^{(2)}$ denotes the contribution second order in $G_\beta$. Greek indices are Lorentz indices and repeated Lorentz indices are summed.

\subsection{Leptonic matrix element}\label{LME}
Focusing on the leptonic part, we observe the following property for fermion fields,
\begin{align}\label{trans-prop}
\overline{e}\argu{x}\mathcal{O}_B\nu_j\argu{x}=\(\overline{e}\argu{x}\mathcal{O}_B\nu_j\argu{x}\)^T=-\nu^T_j\argu{x}\mathcal{O}_B^T\overline{e}^T\argu{x}.
\end{align}
Applying Wick's theorem we can write the leptonic time-ordered product in terms of normal-ordering ($\mathcal{N}$) and contraction. Using the Majorana condition,
\begin{align}
 &\nu=\nu^C\Rightarrow \nu^T=-\overline{\nu}\mathcal{C},
\end{align}
where the charge-conjugated field $\nu^C=\mathcal{C}\overline{\nu}^T$ is defined in terms of the charge-conjugation
matrix $\mathcal{C}$, we can write the leptonic matrix element (LME) in terms of the neutrino loop four-momentum $q^\mu$ for $B\in\{V-A,S+P\}$:
\begin{align}\label{gen-lep-elem}
\matrixel{e_1^-,e_2^-}{\mathscr{T}\left[j^{\rho j}_{V-A}\argu{x_1}j^j_{B}\argu{x_2}\right]}{0}&= i\qqint\frac{e^{-iq\cdot(x_1-x_2)}}{q^2-m_j^2+i\epsilon}\nn\\ 
&\times\matrixel{e_1^-,e_2^-}{\mathcal{N}\left[ 
\overline{e}\argu{x_1}\mathcal{O}_{V-A}(\Gm q_{\mu}+m_j)\mathcal{C}\mathcal{O}_B^T\overline{e}^T\!\!\!\argu{x_2}\right]}{0}
\end{align}
Here $m_j$ is the mass of the $j$th neutrino mass eigenstate and
$\overline e\equiv e^\dagger\gamma^0$. In Eq.~\eqref{gen-lep-elem},
$q^\mu=(q^0,\bm q)$ is the integration variable. After the $q^0$ contour
integral, its positive-energy pole is
$q^0=\omega_j\equiv\sqrt{\bm q^2+m_j^2}$; below we abbreviate $\omega_j$
as $\omega$. The $\epsilon$ in the propagator denominator is the positive
infinitesimal of the Feynman prescription and is unrelated to either
$\varepsilon_{A,j}^{B}$ or the amplitude-level couplings. We use
$\mathcal C^T=-\mathcal C$ and
$\mathcal C(\gamma^\mu)^T\mathcal C^{-1}=-\gamma^\mu$.

The chiral projectors distinguish neutrino-mass and neutrino-momentum
numerators:
\begin{align}
&\mathcal{O}_{V-A}(\Gm q_{\mu}+m_j)\mathcal{C}\mathcal{O}_{V-A}^T=-\Gr\pl(\Gm q_{\mu}+m_j)\pl\Gs\mathcal{C}=-2m_j\Gr\Gs\pr\mathcal{C},\label{v-av-a}\\  
&\mathcal{O}_{V-A}(\Gm q_{\mu}+m_j)\mathcal{C}\mathcal{O}_{S+P}^T =\Gr\pl(\Gm q_{\mu}+m_j)\pr\mathcal{C}=2\Gr\Gm q_{\mu}\pr\mathcal{C},\label{v-as+p}
 \end{align}
Thus the $j_{V-A}j_{V-A}$ contribution is mass dependent, whereas the
scalar contribution is momentum dependent. The projector identities used
above are
\begin{align}\label{projections}
& \prl(\Gm q_{\mu}+m_j)\prl=2m_j\prl,\nn \\ 
& \prl(\Gm q_{\mu}+m_j)\plr=2\Gm q_{\mu}\plr, 
\end{align}

After the Wick expansion, the leptonic matrix element is evaluated using the plane-wave expansion of the electron fields. Integrating the time-dependent factors after performing the neutrino-energy integral \cite{bilenky2010introduction}, we arrive at the $\mathcal S$-matrix elements for the three terms of Eq.~\eqref{0nmel}:
\begin{align}
&\mel**{f}{\mathcal{S}^{(2)}_{0\nu}}{i}_{V-A}^{V-A}=\frac{-i}{(2\pi)^3}\frac{G^2_{\beta}}{2} \effmass\intinfdds\momintomega\Big(2\pi\delta(E_i-E_f-\varepsilon_1-\varepsilon_2)\Big)\sum_a \nn\\
&\qquad\qquad\times\Bigg[\frac{\overline{\Psi}_{s_1}\argu{\varepsilon_1,\bm{x}_1}\Gr\Gs\pr\Psi^C_{s_2}\argu{\varepsilon_2,\bm{x}_2}}{\omega+E_a-E_i+\varepsilon_2}-\frac{\overline{\Psi}_{s_2}\argu{\varepsilon_2,\bm{x}_1}\Gr\Gs\pr\Psi^C_{s_1}\argu{\varepsilon_1,\bm{x}_2}}{\omega+E_a-E_i+\varepsilon_1}\Bigg]\nn\\ 
&\qquad\qquad\quad\times\mel**{N_f}{J^{\dagger}_{V-A,\rho}(\bm{x_1})}{N_a}\mel**{N_a}{J^{\dagger}_{V-A,\sigma}(\bm{x_2})}{N_i},\label{Sv-av-a}\\
&\mel**{f}{\mathcal{S}^{(2)}_{0\nu}}{i}_{S\pm P}^{S+P}=\frac{i}{(2\pi)^3}\frac{G^2_{\beta}}{2}\epsilon_{S\pm P}^{S+P}\intinfdds\momintomega\Big(2\pi \delta(E_i-E_f-\varepsilon_1-\varepsilon_2)\Big)\sum_a q_{\mu}\nn\\
&\qquad\qquad\times\Bigg\{\Bigg[\frac{\overline{\Psi}_{s_1}\argu{\varepsilon_1,\bm{x}_1}\Gr\Gm\pr\Psi^C_{s_2}\argu{\varepsilon_2,\bm{x}_2}}{\omega+E_a-E_i+\varepsilon_2}-\frac{\overline{\Psi}_{s_2}\argu{\varepsilon_2,\bm{x}_1}\Gr\Gm\pr\Psi^C_{s_1}\argu{\varepsilon_1,\bm{x}_2}}{\omega+E_a-E_i+\varepsilon_1}\Bigg]\nn\\ &\qquad\qquad\qquad\qquad\times\mel**{N_f}{J^{\dagger}_{V-A,\rho}(\bm{x_1})}{N_a}\mel**{N_a}{J^{\dagger}_{S\pm P}(\bm{x_2})}{N_i}\nn\\
&\qquad\qquad\quad-\Bigg[\frac{\overline{\Psi}_{s_1}\argu{\varepsilon_1,\bm{x}_1}\Gm\Gr\pr\Psi^C_{s_2}\argu{\varepsilon_2,\bm{x}_2}}{\omega+E_a-E_i+\varepsilon_2}-\frac{\overline{\Psi}_{s_2}\argu{\varepsilon_2,\bm{x}_1}\Gm\Gr\pr\Psi^C_{s_1}\argu{\varepsilon_1,\bm{x}_2}}{\omega+E_a-E_i+\varepsilon_1}\Bigg]\nn\\ 
&\qquad\qquad\quad\times\mel**{N_f}{J^{\dagger}_{S\pm P}(\bm{x_1})}{N_a}\mel**{N_a}{J^{\dagger}_{V-A,\rho}(\bm{x_2})}{N_i}\!\!\!\Bigg\}.\label{Ss-ps+p}
\end{align}
Here we have redefined the LNV parameters at the amplitude level as
\begin{align}\label{redefine_eff_param}
 \sum_j U_{ej}\varepsilon_{S\pm P, j}^{S+P}=\epsilon_{S\pm P}^{S+P}.
 \end{align}
$\effmass$ is the effective neutrino mass,
 \begin{align}\label{m_eff}
& \sum_j U_{ej}^2 m_j=\effmass.
 \end{align}
A complete set of intermediate nuclear states $\sum_a\ket{N_a}\bra{N_a}$ is inserted for the hadronic matrix element, with $E_a$ the energy of state $\ket{N_a}$. The initial $\ket{N_i}$ and final $\ket{N_f}$ nuclear states have energies $E_i$ and $E_f$, respectively. Unlike the scalar terms in Eq.~\eqref{Ss-ps+p}, the mass-dependent case in Eq.~\eqref{Sv-av-a} contains one current ordering rather than two. The full $\Smat$ element for the mass and scalar terms of Eq.~\eqref{Ham-longrange} is then
\begin{align}\label{fullSmatrix}
\mel**{f}{\mathcal{S}^{(2)}_{0\nu}}{i}=&\mel**{f}{\mathcal{S}^{(2)}_{0\nu}}{i}_{V-A}^{V-A}+\mel**{f}{\mathcal{S}^{(2)}_{0\nu}}{i}_{S-P}^{S+P}+\mel**{f}{\mathcal{S}^{(2)}_{0\nu}}{i}_{S+P}^{S+P},
\end{align}
with Eqs.~\eqref{Sv-av-a}--\eqref{Ss-ps+p} giving the exact expressions.
Here $\Psi$ denotes a Coulomb-distorted electron wavefunction; its
$s_{1/2}$ and $p_{1/2}$ components are given in
Appendix~\ref{elec-wavefunc}.  The scalar amplitude is reduced to its
energy- and momentum-numerator parts in Appendix~\ref{scalarterms}.
The labels $\varepsilon_{1,2}$ and $s_{1,2}$ denote the total energies and spin projections of the two outgoing electrons, respectively, $\bm x_{1,2}$ are the spatial parts of $x_{1,2}$, $\delta$ is the Dirac delta enforcing energy conservation, and $\Psi_s^C\equiv\mathcal C\overline{\Psi}_s^T$ is the charge-conjugated electron wavefunction.

\subsection{Nuclear matrix elements}\label{NME}
For the calculation of hadronic matrix elements, the color-singlet quark
current $\overline{u}\mathcal O_A d$ must first be matched to the nucleon
level and then embedded in the nucleus.  For the free transition
$n\rightarrow p+e^-+\overline\nu_e$, the relevant one-nucleon quantity is
$\langle p|\overline{u}\mathcal O_A d|n\rangle$.  Protons and neutrons are
baryonic bound states, so their quark substructure is encoded in nucleon form
factors.  In terms of the nucleon isodoublet $N=(p\;n)^T$ and the isospin
raising operator $\tau^+$, we write
\begin{align}
 \langle p(p_p)|\overline u\mathcal O_A d|n(p_n)\rangle
 =\overline N(p_p)\tau^+\mathcal J_A(k)N(p_n),
\end{align}
where $p_n^\mu$ and $p_p^\mu$ are, respectively, the initial-neutron and
final-proton four-momenta. We use
$\tau^+=(\tau^1+i\tau^2)/2$, so that $\tau^+|n\rangle=|p\rangle$, and
throughout adopt
\begin{align}
 k^\mu\equiv p_n^\mu-p_p^\mu,
 \qquad K^\mu\equiv p_n^\mu+p_p^\mu,
 \qquad Q^2\equiv-k^2\geq0,
 \label{momentum-transfer}
\end{align}
so $k^\mu$ is the momentum transferred from the nucleon to the leptonic
system at a given vertex.  With the electron and virtual-neutrino momenta
taken as outgoing from that vertex, four-momentum conservation gives
$k^\mu=p_e^\mu+q^\mu$, where $q^\mu$ is the internal-neutrino momentum
used in Sec.~\ref{LME}.  Since
$|\bm p_e|\ll|\bm q|$ in the nucleus, one may nevertheless use
$\bm k\simeq\bm q$ in the leading nuclear operators.

For the scalar--pseudoscalar and vector--axial-vector quark currents, our
dimensionless form-factor convention is
\begin{align}
 \langle p(p_p)|\overline u(1\pm\gamma_5)d|n(p_n)\rangle
 &=\overline N(p_p)\tau^+
 \left[F_S(Q^2)\pm F_{PS}(Q^2)\gamma_5\right]N(p_n),
 \label{S+/-Pnucl}\\
 \langle p(p_p)|\overline u\gamma^\mu(1-\gamma_5)d|n(p_n)\rangle
 &=\overline N(p_p)\tau^+\Bigg[
 g_V(Q^2)\gamma^\mu
 -i\frac{g_M(Q^2)}{2m_N}\sigma^{\mu\nu}k_\nu
 \nonumber\\[-2pt]
 &\qquad\qquad
 -g_A(Q^2)\gamma^\mu\gamma_5
 +\frac{g_P(Q^2)}{2m_N}k^\mu\gamma_5
 \Bigg]N(p_n),
 \label{V+/-Anucl}
\end{align}
where $m_p$ and $m_n$ are the proton and neutron masses,
$m_N=(m_p+m_n)/2$ is their average, and
$\sigma^{\mu\nu}=\tfrac{i}{2}[\gamma^\mu,\gamma^\nu]$.  The factors
$1/(2m_N)$ make both the weak-magnetism form factor $g_M$ and the induced
axial-pseudoscalar form factor $g_P$ dimensionless.  The latter must not be
confused with the pseudoscalar-density form factor $F_{PS}$ in
Eq.~\eqref{S+/-Pnucl}.
The momentum argument $Q^2$ is suppressed below when no ambiguity arises;
$g_V$ and $g_A$ without arguments in overall normalizations denote their
reference values at $Q^2=0$.

At zero momentum transfer, the conserved-vector-current (CVC) relation gives
$g_V(0)=1$ in the isospin limit.  We
define $g_M$ as the isovector Pauli, or anomalous magnetic, form factor,
\begin{align}
 g_M(0)=\kappa_p-\kappa_n\simeq3.706,
 \label{weak-magnetism}
\end{align}
where $\mu_p$ and $\mu_n$ denote the dimensionless total magnetic-moment
coefficients in nuclear magnetons, while
$\kappa_p=\mu_p-1$ and $\kappa_n=\mu_n$ are the corresponding proton and
neutron anomalous coefficients \cite{Hayen:2021,PDG2024}. Thus the
coefficient of the spin--magnetization recoil operator is
$1+g_M/g_V\simeq4.706$, the sum of the Dirac and Pauli contributions.  This
dimensionless $g_M$ is related to the dimensionful weak-magnetism convention
of Doi, Kotani, and Takasugi by
$g_W^{\rm DK}/g_V=-g_M/(2m_Ng_V)$, so
$[g_W^{\rm DK}]=M^{-1}$ \cite{Doi+Kotani1985}.

For reference, the traditional finite-size parametrization used in the
present analytic reduction is
\begin{align}
 g_X(Q^2)&=\frac{g_X(0)}{(1+Q^2/M_X^2)^2},
 &&X=V,A,M,\nonumber\\
 F_S(Q^2)&=\frac{g_S}{(1+Q^2/M_S^2)^2},\nonumber\\
 F_{PS}(Q^2)&=\frac{g_{PS}}{(1+Q^2/M_{PS}^2)^2}
 \frac{m_\pi^2}{Q^2+m_\pi^2},\nonumber\\
 g_P(Q^2)&=\frac{4m_N^2g_A(0)}{Q^2+m_\pi^2}
 \frac{1-m_\pi^2/M_A^2}{(1+Q^2/M_A^2)^2}.
 \label{form-factor-model}
\end{align}
Here $m_\pi$ is the charged-pion mass.  The parameters $M_V$, $M_A$, $M_M$,
$M_S$, and $M_{PS}$ are, respectively, the dipole mass scales governing the
momentum dependence of the vector, axial-vector, weak-magnetism,
scalar-density, and pseudoscalar-density form factors.  The pion-pole
dependence of $F_{PS}$ and $g_P$ is displayed separately through $m_\pi$ in
Eq.~\eqref{form-factor-model}.  Common benchmark values are
$M_V=M_M\simeq0.84$ GeV, $M_A\simeq1.0$ GeV, $g_A(0)\simeq1.2754$, and
$M_S,M_{PS}\sim M_V$.  The scalar and pseudoscalar charges are
renormalization-scheme and scale dependent; a frequently used matched input
is $g_S=1.02(11)$ and $g_{PS}=349(9)$ in the $\overline{\rm MS}$ scheme at
the renormalization scale $\mu=2$ GeV
\cite{GonzalezAlonso:2014, GonzalezAlonso:2019}.  Accordingly,
$F_S$, $F_{PS}$ and their EFT coefficients must always be evaluated in the
same scheme at the same scale.  Equation~\eqref{form-factor-model} is a
transparent dipole/pion-pole working parametrization, not a claim that a
single dipole is the unique modern representation of the measured form
factors.  The analytic formulas below remain valid if it is replaced by
updated empirical or lattice quantum chromodynamics (QCD) form factors,
provided their momentum
dependence is retained consistently inside the neutrino-potential integrals
\cite{Cirigliano:2018yza}.

The nucleon currents are next expanded nonrelativistically.  Since typical
nuclear momenta are small compared with $m_N$, we retain terms through first
order in nucleon recoil and use the single-nucleon impulse approximation.
The scalar and vector nuclear currents then take the form
\cite{ericson1988pions,PhysRevD.105.099902}
\begin{align}
 J^{\dagger}_{S\pm P}(\bm x)
 &=\sum_{n=1}^A\tau_n^+\delta^{(3)}(\bm x-\bm r_n)
 \left[F_S\mathbbm{1}_n\pm F_{PS}B_n\right],
 \label{Js+/-pnr}\\
 J_{V-A}^{\mu\dagger}(\bm x)
 &=\sum_{n=1}^A\tau_n^+\delta^{(3)}(\bm x-\bm r_n)
 \left[(g_V\mathbbm{1}_n-g_AC_n)g^{\mu0}
 +(g_A\sigma_n^i-g_VD_n^i-g_AP_n^i)g^{\mu i}\right].
 \label{Jv-anr}
\end{align}
Here $A$ is the nuclear mass number, $\bm r_n$ and $\bm\sigma_n$ are the
position and Pauli-spin operators of nucleon $n$, $\tau_n^+$ converts a
neutron into a proton, and $\mathbbm{1}_n$ is the identity in its spin
space. Here $\delta^{(3)}$ is the three-dimensional Dirac delta. For an
individual nucleon let
\begin{align}
 \bm k_n\equiv\bm p_n-\bm p_n',\qquad
 k_n^0\equiv E_n-E_n',\qquad
 \bm K_n\equiv\bm p_n+\bm p_n',
\end{align}
where unprimed and primed quantities refer to the initial and final nucleon.
Using the notation introduced above, the dimensionless pseudoscalar and recoil operators
are
\begin{align}
 B_n&=\frac{\bm\sigma_n\cdot\bm k_n}{2m_N},
 \label{B}\\
 C_n&=\frac{\bm K_n\cdot\bm\sigma_n}{2m_N}
 -\frac{g_P}{g_A}\frac{k_n^0\,\bm\sigma_n\cdot\bm k_n}{4m_N^2},
 \label{C}\\
 \bm D_n&=\frac{\bm K_n}{2m_N}
 -\left(1+\frac{g_M}{g_V}\right)
 \frac{i\bm\sigma_n\times\bm k_n}{2m_N},
 \label{D}\\
 \bm P_n&=\frac{g_P}{g_A}
 \frac{\bm k_n(\bm\sigma_n\cdot\bm k_n)}{4m_N^2}.
 \label{P}
\end{align}
Equations~\eqref{B}--\eqref{P} follow the
momentum direction fixed in Eq.~\eqref{momentum-transfer}.  The full one-body
current contains $\bm P_n$ as displayed.  In the dominant scalar-recoil
subset used later, contributions generated by $\bm P_n$, products of two
recoil operators such as $B_nC_m$ and $B_n\bm D_m$, and the subleading
$C_{4R}^A$ term are omitted explicitly; this is a stated working truncation,
not an operator identity \cite{PhysRevD.105.099902}.

The presence of intermediate nuclear states poses a computationally
challenging problem, which is considerably simplified by the closure
approximation.  The dominant neutrino three-momentum satisfies
$|\bm q|\sim1/r_{nm}\sim100$--$200$ MeV for an internucleon separation
$r_{nm}\equiv|\bm r_n-\bm r_m|\sim1$--$2$ fm, whereas the spread of the intermediate-state
excitation energies is typically of order $10$ MeV.  We therefore replace
$E_a$ by an average closure energy $\langle E_a\rangle$ and use completeness
to perform the sum over intermediate nuclear states
\cite{Senkov:2013gso}.  The contraction of these one-body currents into the
two-nucleon scalar operators is given in Appendix~\ref{scalarterms}.

\section{Decay rate}\label{decay-rate-section}
The differential decay rate for $\NDBD$ is
\begin{align}\label{diffdecayrate}
 d\Gamma_{0\nu}={}&2\pi\sum_{s_1,s_2}|\mathcal R_{0\nu}|^2
 \delta(\varepsilon_1+\varepsilon_2+E_f-E_i)
 \frac{\dd[3]{\bm p_1}}{(2\pi)^3}
 \frac{\dd[3]{\bm p_2}}{(2\pi)^3}.
\end{align}
Here $\mathcal R_{0\nu}$ is the transition amplitude after removing the
overall energy-conserving delta function, $\bm p_{1,2}$,
$\varepsilon_{1,2}$, and $s_{1,2}$ are the momenta, total energies, and spin
projections of the two outgoing electrons, and $E_i$ and $E_f$ are the
initial- and final-nuclear energies.  The spin sum runs over $s_1$ and $s_2$.
For the mechanisms retained in this work,
\begin{align}
 \mathcal R_{0\nu}=\mathcal R^m_{0\nu}+\mathcal R^{SP}_{0\nu},
 \label{Rfull}
\end{align}
where $\mathcal R^m_{0\nu}$ is the standard light-Majorana-neutrino mass
mechanism and $\mathcal R^{SP}_{0\nu}$ contains the two long-range scalar
interactions. The symbol $m_e$ below denotes the electron mass.

\subsection{Standard mass mechanism}
The standard mass mechanism has been studied extensively together with the
$V\pm A$ interactions \cite{Doi+Kotani1985,PhysRevD.76.093009}.  Following
the notation of the preceding section, its closure-approximated amplitude for
a $0^+\to0^+$ transition can be written as
\begin{align}
 \mathcal R^m_{0\nu}&=C_{0\nu}g_A^2Z_1^X
 E_+^{s_{1/2}s_{1/2}},
 \label{ampVA2}\\
 C_{0\nu}&=\frac{G_\beta^2m_e}{4\sqrt2\pi R},
 \qquad [C_{0\nu}]=M^{-2},
 \label{C0nu-amplitude}\\
 Z_1^X&=\eta_\nu M_{GT}^{0\nu}(\chi_F-1),
 \qquad
 \eta_\nu\equiv\frac{\effmass}{m_e}.
 \label{Mmfinal}
\end{align}
The coefficient $C_{0\nu}$ normalizes the amplitude and must be distinguished
from the rate normalization introduced when Eq.~\eqref{diffdecayrate} is
integrated.  The effective Majorana mass $\effmass$ was defined in
Eq.~\eqref{m_eff}; hence $\eta_\nu$ is dimensionless.

The dimensionless Fermi and Gamow--Teller NME and their ratio are
\begin{align}
 M_F^{0\nu}&=\dblmatel{0_f^+}{
 \sum_{n\ne m}\tau_n^+\tau_m^+h_+(r_{nm})}{0_i^+},
 \label{MF0nu}\\
 M_{GT}^{0\nu}&=\dblmatel{0_f^+}{
 \sum_{n\ne m}\tau_n^+\tau_m^+h_+(r_{nm})
 \bm\sigma_n\cdot\bm\sigma_m}{0_i^+},
 \label{MGT0nu}\\
 \chi_F&=\left(\frac{g_V}{g_A}\right)^2
 \frac{M_F^{0\nu}}{M_{GT}^{0\nu}}.
 \label{chiF}
\end{align}
Thus $Z_1^X$ is the mass-mechanism NME combination, including the effective
mass parameter.  The superscript $X$ follows the conventional notation for
the standard mechanism.  The double bars denote reduced nuclear matrix
elements between the initial and final $0^+$ states, and $n\ne m$ labels the
two distinct decaying nucleons.

The closure neutrino potential appearing in Eqs.~\eqref{MF0nu} and
\eqref{MGT0nu} is
\begin{align}
 h_+(r_{nm},\langle E_a\rangle)
 &=\frac{R}{2}\left[H_2(r_{nm},\langle E_a\rangle)
 +H_1(r_{nm},\langle E_a\rangle)\right],
 \label{hplus}\\
 H_{1(2)}(r_{nm},\langle E_a\rangle)
 &=\frac{1}{2\pi^2}\int\frac{\dd[3]{\bm q}}{\omega}
 \frac{e^{i\bm q\cdot\bm r_{nm}}}
 {\omega+\langle E_a\rangle-\tfrac12(E_i+E_f)
 \pm\tfrac12(\varepsilon_1-\varepsilon_2)},
 \qquad \omega\simeq|\bm q|.
 \label{nu-pot1}
\end{align}
The upper and lower signs in Eq.~\eqref{nu-pot1} define $H_1$ and $H_2$,
respectively. Here $\bm r_{nm}=\bm r_n-\bm r_m$,
$r_{nm}=|\bm r_{nm}|$, and $R=r_0A^{1/3}$ is the conventional nuclear
radius, with $r_0\simeq1.2$ fm. The factor $R$ makes $h_+$, and consequently the NME in
Eqs.~\eqref{MF0nu} and \eqref{MGT0nu}, dimensionless.  The momentum $\bm q$
is the virtual-neutrino momentum, not the one-nucleon momentum transfer
$\bm k$ defined in Eq.~\eqref{momentum-transfer}. Below, closure-energy and
electron-energy arguments of the neutrino potentials are suppressed when
they are unchanged.

The leading electron bilinear is
\begin{align}
 E_+^{s_{1/2}s_{1/2}}
 =\overline{\Psi_{s_1}^{s_{1/2}}}(\varepsilon_1,R)
 (1+\gamma_5)
 \Psi_{s_2}^{s_{1/2}C}(\varepsilon_2,R),
 \label{E+ss}
\end{align}
where $\Psi^{s_{1/2}}_s$ is a Coulomb-distorted outgoing-electron wavefunction
and $\Psi^C=\mathcal C\overline{\Psi}^{T}$.  The slowly varying radial
functions are evaluated at $r=R$ in the usual surface-factorization
approximation described in Appendix~\ref{elec-wavefunc}.

\subsection{Scalar-type \texorpdfstring{$S\pm P$}{S plus/minus P} terms}
We now consider the two scalar interactions with the common $S+P$ leptonic
current.  Their detailed amplitude reduction is given in
Appendix~\ref{scalarterms}.  For compactness we define
\begin{align}
 \epsilon_-&\equiv\epsilon_{S-P}^{S+P},
 &\epsilon_+&\equiv\epsilon_{S+P}^{S+P},
 \label{scalar-coupling-short}\\
 \widetilde J_{SP}^\dagger
 &=\epsilon_+J_{S+P}^\dagger+\epsilon_-J_{S-P}^\dagger.
 \label{combined-scalar-current}
\end{align}
The upper label specifies the leptonic current and the lower label the
hadronic current.  Both $\epsilon_-$ and $\epsilon_+$ are the amplitude-level
couplings defined by Eq.~\eqref{redefine_eff_param}; they may be complex.
Their scalar and pseudoscalar nucleon combinations are
\begin{align}
 G_V^0&\equiv\frac{g_V}{g_A},
 &\varepsilon_S'&\equiv\frac{F_S(Q^2)}{g_A}(\epsilon_++\epsilon_-),
 &\varepsilon_P'&\equiv\frac{F_{PS}(Q^2)}{g_A}(\epsilon_+-\epsilon_-).
 \label{SP-comp-coupl}
\end{align}
The distinction between the amplitude-level $\epsilon_\pm$ and
the Hamiltonian coefficients $\varepsilon_{A,j}^{B}$ is the one established
in Eq.~\eqref{redefine_eff_param}.

Before displaying the scalar amplitude, we define all electron, neutrino, and
nuclear quantities entering it.  Let
\begin{align}
 \Psi_1(\bm x_\alpha)&\equiv
 \Psi_{s_1}(\varepsilon_1,\bm x_\alpha),
 &\Psi_2(\bm x_\alpha)&\equiv
 \Psi_{s_2}(\varepsilon_2,\bm x_\alpha),
 &\Delta_\alpha&\equiv E_a-E_i+\varepsilon_\alpha,
 \qquad \alpha=1,2.
\end{align}
The label $\alpha$ distinguishes the two time orderings.  With
$\bm r=\bm x_1-\bm x_2$, the energy- and
momentum-numerator neutrino potentials are
\begin{align}
 H_{\omega\alpha}(\bm r,E_a)
 &\equiv\frac{1}{2\pi^2}\int\frac{\dd[3]{\bm q}}{\omega}
 \frac{\omega}{\omega+\Delta_\alpha}e^{i\bm q\cdot\bm r},
 \label{neu-pot-energy}\\
 H_{q\alpha}^{l}(\bm r,E_a)
 &\equiv\frac{1}{2\pi^2}\int\frac{\dd[3]{\bm q}}{\omega}
 \frac{q^l}{\omega+\Delta_\alpha}e^{i\bm q\cdot\bm r}.
 \label{neu-pot-mom}
\end{align}
Here $l$ is a spatial Cartesian index, and repeated spatial indices are
summed.  We use the exchange-even and exchange-odd electron
bilinears
\begin{align}
 E_\pm&\equiv\frac12\left[
 \overline\Psi_1(\bm x_1)(1+\gamma_5)\Psi_2^C(\bm x_2)
 \pm\overline\Psi_1(\bm x_2)(1+\gamma_5)\Psi_2^C(\bm x_1)
 \right],
 \label{Epm-definition}\\
 E_\pm^l&\equiv\frac12\left[
 \overline\Psi_1(\bm x_1)\gamma^l\gamma^0(1+\gamma_5)
 \Psi_2^C(\bm x_2)
 \pm\overline\Psi_1(\bm x_2)\gamma^l\gamma^0(1+\gamma_5)
 \Psi_2^C(\bm x_1)
 \right].
 \label{Epml-definition}
\end{align}

For a nucleon pair $(n,m)$, the combinations of the one-body operators
defined in Eqs.~\eqref{B}--\eqref{D} are
\begin{align}
 B_\pm&=B_n\mathbbm1_m\pm\mathbbm1_nB_m,
 &C_\pm&=C_n\mathbbm1_m\pm\mathbbm1_nC_m,
 \nonumber\\
 \sigma_\pm^l&=\sigma_n^l\mathbbm1_m\pm\mathbbm1_n\sigma_m^l,
 &D_\pm^l&=D_n^l\mathbbm1_m\pm\mathbbm1_nD_m^l,
 \nonumber\\
 B_{\sigma\pm}^l&=\sigma_n^lB_m\pm B_n\sigma_m^l.
 \label{pair-operators}
\end{align}
Here $\mathbbm1_n$ is the identity in the spin space of nucleon $n$.  The
nuclear operators used below are then
\begin{align}
 A_2&=2G_V^0\varepsilon_S'\mathbbm1_n\mathbbm1_m,
 &A_{2R}&=-\varepsilon_S'C_++G_V^0\varepsilon_P'B_+,
 \nonumber\\
 A_4^l&=\varepsilon_S'\sigma_+^l,
 &A_{4R}^l&=\varepsilon_P'B_{\sigma+}^l
             -G_V^0\varepsilon_S'D_+^l,
 \nonumber\\
 A_5^{lk}&=i\epsilon^{ilk}\varepsilon_S'\sigma_+^i,
 &A_{5R}^{lk}&=i\epsilon^{ilk}
 \left(\varepsilon_P'B_{\sigma+}^i-G_V^0\varepsilon_S'D_+^i\right),
 \nonumber\\
 B_{2R}&=-G_V^0\varepsilon_P'B_- -\varepsilon_S'C_-,
 \nonumber\\
 B_4^l&=\varepsilon_S'\sigma_-^l,
 &B_{4R}^l&=\varepsilon_P'B_{\sigma-}^l
             -G_V^0\varepsilon_S'D_-^l.
 \label{A_B-SP}
\end{align}
The symbol $\epsilon^{ilk}$ is the three-dimensional Levi--Civita symbol,
with $\epsilon^{123}=+1$, and
the subscript $R$ identifies recoil-order contributions.  Products containing two recoil operators, such
as $B_nC_m$ and $B_n\bm D_m$, are beyond the retained order.

The intermediate-state transition density acts on a factorized pair
operator as
\begin{align}
T_a[\mathcal O_n\mathcal O_m]\equiv g_A^2\sum_{n\ne m}
&\matrixel{N_f}{
\tau_n^+\delta^{(3)}(\bm x_1-\bm r_n)\mathcal O_n}{N_a}
\nonumber\\
\times{}&
\matrixel{N_a}{
\tau_m^+\delta^{(3)}(\bm x_2-\bm r_m)\mathcal O_m}{N_i},
 \label{Ta}
\end{align}
where $|N_a\rangle$ is an intermediate nuclear state.  The notation
$T_a\mathcal O$ below abbreviates the linear action $T_a[\mathcal O]$.
With these definitions, the scalar amplitude takes the form
\begin{align}
 \mathcal R_{0\nu}^{SP}
 &=C_{0\nu}\sum_a\left[
 \{M_{\bm q}^{SP}\}_{n}+\{M_{\bm q}^{SP}\}_{c}\right],
 \label{Full-S-scalar}\\
 \{M_{\bm q}^{SP}\}_{n}
 &=\frac{R}{2m_e}\int\dd[3]{\bm x_1}\dd[3]{\bm x_2}\,T_a
 \Bigg\{(H_{\omega2}-H_{\omega1})
 \left[-(A_4^l+A_{4R}^l)E_+^l+B_{2R}E_-\right]
 \nonumber\\
 &\hspace{2.4cm}+(H_{q2}^l+H_{q1}^l)
 \left[-(A_2+A_{2R})E_-^l
 +(A_5^{lk}+A_{5R}^{lk})E_-^k
 -(B_4^l+B_{4R}^l)E_+\right]\Bigg\},
 \label{MqSPn}\\
 \{M_{\bm q}^{SP}\}_{c}
 &=\frac{R}{2m_e}\int\dd[3]{\bm x_1}\dd[3]{\bm x_2}\,T_a
 \Bigg\{(H_{\omega2}+H_{\omega1})
 \left[-(A_4^l+A_{4R}^l)E_-^l+B_{2R}E_+\right]
 \nonumber\\
 &\hspace{2.4cm}+(H_{q2}^l-H_{q1}^l)
 \left[-(A_2+A_{2R})E_+^l
 +(A_5^{lk}+A_{5R}^{lk})E_+^k
 -(B_4^l+B_{4R}^l)E_-\right]\Bigg\}.
 \label{MqSPc}
\end{align}
In this notation, $\{M_{\bm q}^{SP}\}_n$ is exchange even and
$\{M_{\bm q}^{SP}\}_c$ is exchange odd.  Under $n\leftrightarrow m$ together
with $\bm x_1\leftrightarrow\bm x_2$, the complete pair sum is symmetric after
$E_a\to\langle E_a\rangle$.  Consequently, Eq.~\eqref{MqSPc} vanishes in
closure, while Eq.~\eqref{MqSPn} survives.

For a $0^+\to0^+$ transition, parity and angular-momentum selection rules
further restrict the operators that can accompany the $s_{1/2}$ and
$p_{1/2}$ electron partial waves defined in
Appendix~\ref{elec-wavefunc}. In the following table, $r^l$ denotes the
spatial coordinate supplied by the first-order $p_{1/2}$ electron-wave
expansion; after the two vertices are combined it becomes a component of
$\bm r_{nm}$ or $\bm r_{nm+}$, as shown explicitly below. The resulting
classification is
\begin{table}[htp]
\centering
\begin{tabular}{lccc}
\hline
Nuclear operator & Parity & $s_{1/2}$--$s_{1/2}$
& $s_{1/2}$--$p_{1/2}$\\
\hline
$A_4^l$                 & even & $0^+\to0^+$ & --\\
$r^lB_{4R}^l$           & even & $0^+\to0^+$ & --\\
$r^lA_{2R}$             & even & $0^+\to0^+$ & --\\
$r^lA_{5R}^{lk}$        & even & $0^+\to0^+$ & --\\
$A_{4R}^l$              & odd  & -- & $0^+\to0^+$\\
$B_{2R}$                & odd  & -- & --\\
$r^lB_4^l$              & odd  & -- & --\\
$r^lA_2$                & odd  & -- & $0^+\to0^+$\\
$r^lA_5^{lk}$           & odd  & -- & $0^+\to0^+$\\
\hline
\end{tabular}
\caption{Parity and electron-partial-wave selection rules for the nuclear
operators contributing to a $0^+\to0^+$ transition.  A dash denotes a
vanishing contribution at the retained order.}
\label{Mq-0to0-tab}
\end{table}

For two $s_{1/2}$ electrons, scalar angular recoupling leaves
\begin{align}
 \sum_a\{M_{\bm q}^{SP}\}_{n}^{s_{1/2}s_{1/2}}
 &=-g_A^2\frac{2}{m_eR}C_{4R}^{B}
 E_+^{s_{1/2}s_{1/2}},
 \label{MqnSP-ss}\\
 C_{4R}^{B}
 &=\dblmatel{0_f^+}{\sum_{n\ne m}\tau_n^+\tau_m^+
 \frac{iR}{2r_{nm}}h_+'\,\widehat{\bm r}_{nm}\cdot\bm B_{4R}}
 {0_i^+},
 \label{C4RB}
\end{align}
where $\bm B_{4R}$ is the vector with components $B_{4R}^l$ and
$\widehat{\bm r}_{nm}=\bm r_{nm}/r_{nm}$.

For one $s_{1/2}$ and one $p_{1/2}$ electron, only the odd-parity operators
that can be coupled to an overall scalar survive.  The result is
\begin{align}
 \sum_a\{M_{\bm q}^{SP}\}_{n}^{s_{1/2}p_{1/2}}
 =g_A^2\left[
 \frac{\varepsilon_{12}}{4m_e}C_{4R}^{A}
 \widetilde E_+^{s_{1/2}p_{1/2}}
 -\frac{1}{2m_eR}(C_2^A-C_5^A)
 \widetilde E_-^{s_{1/2}p_{1/2}}
 \right],
 \label{MqnSP-sp}
\end{align}
where $\varepsilon_{12}\equiv\varepsilon_1-\varepsilon_2$.  The associated
NME combinations are
\begin{align}
 C_{4R}^{A}
 &=\dblmatel{0_f^+}{\sum_{n\ne m}\tau_n^+\tau_m^+
 \frac{i}{2r_{nm+}}h_{0\omega}\,
 \bm r_{nm+}\cdot\bm A_{4R}}{0_i^+},
 \label{C4RA}\\
 C_2^A
 &=\dblmatel{0_f^+}{\sum_{n\ne m}\tau_n^+\tau_m^+
 h_+'A_2}{0_i^+},
 \label{C2A}\\
 C_5^A
 &=\dblmatel{0_f^+}{\sum_{n\ne m}\tau_n^+\tau_m^+
 h_+'\widehat r_{nm}^{l}\widehat r_{nm}^{k}A_5^{lk}}{0_i^+}.
 \label{C5A}
\end{align}
Here $\bm r_{nm+}=\bm r_n+\bm r_m$,
$r_{nm+}=|\bm r_{nm+}|$,
$\widehat{\bm r}_{nm+}=\bm r_{nm+}/r_{nm+}$, and $\bm A_{4R}$ has
components $A_{4R}^l$.
The reduced electron bilinears $\widetilde E_\pm^{s_{1/2}p_{1/2}}$ are
defined by factoring the leading coordinate dependence from
$E_\pm^l$:
\begin{align}
 (E_+^l)^{s_{1/2}p_{1/2}}
 &\simeq\frac{ir_{nm+}}{R}\widehat r_{nm+}^{l}
 \widetilde E_+^{s_{1/2}p_{1/2}},
 \nonumber\\
 (E_-^l)^{s_{1/2}p_{1/2}}
 &\simeq\frac{ir_{nm}}{2R}\widehat r_{nm}^{l}
 \widetilde E_-^{s_{1/2}p_{1/2}}.
 \label{E-tilde}
\end{align}
Both assignments--electron 1 in the $p_{1/2}$ wave and electron 2 in the
$s_{1/2}$ wave, and the exchanged assignment--are included in these reduced
bilinears.

The two dimensionless neutrino potentials required in
Eqs.~\eqref{C4RB}--\eqref{C5A} are defined by
\begin{align}
 h_{0\omega}
 &\equiv\frac{R}{\varepsilon_{12}}(H_{\omega2}-H_{\omega1}),
 \nonumber\\
 h_+'\widehat r_{nm}^{l}
 &\equiv\frac{Rr_{nm}}{2i}(H_{q2}^{l}+H_{q1}^{l}).
 \label{H0w-H+}
\end{align}
Finally, $C_5^A$ vanishes identically because $A_5^{lk}$ is antisymmetric in
$l$ and $k$, whereas $\widehat r_{nm}^{l}\widehat r_{nm}^{k}$ is
symmetric:
\begin{align}
 \widehat r_{nm}^{l}\widehat r_{nm}^{k}A_5^{lk}
 =i\varepsilon_S'\bm\sigma_+\cdot
 (\widehat{\bm r}_{nm}\times\widehat{\bm r}_{nm})=0.
\end{align}
Thus the operator reduction leaves $C_{4R}^{B}$, $C_{4R}^{A}$, and $C_2^A$ before
the stated truncation.  In the dominant scalar-recoil subset used for the
compact rate, the subleading $C_{4R}^{A}$ term is omitted.  Because
$\varepsilon_S'$ and $\varepsilon_P'$ contain both $\epsilon_-$ and
$\epsilon_+$ coherently, the amplitude above already retains the two scalar
mechanisms simultaneously; their mutual interference appears when
$|\mathcal R_{0\nu}|^2$ is formed in the next section.

\section{Half-life formula for scalar-type terms}\label{half-life-section}
We now form the spin-summed square of the total amplitude in
Eq.~\eqref{Rfull}.  The electron spin sums and the associated kinematic
functions are derived in Appendix~\ref{scalar-spin-sums}.  We define
\begin{align}
a_{0\nu}^{\rm rate}
&=\frac{(G_\beta g_A)^4m_e^9}{64\pi^5},
\qquad [a_{0\nu}^{\rm rate}]=M,
\label{a0nu-rate}\\
\dd\Omega_{0\nu}
&=m_e^{-5}p_1p_2\varepsilon_1\varepsilon_2
\delta(\varepsilon_1+\varepsilon_2+E_f-E_i)
\dd{\varepsilon_1}\dd{\varepsilon_2}
\dd{(\widehat{\bm p}_1\cdot\widehat{\bm p}_2)}.
\label{phase-space-measure}
\end{align}
Here $p_i=|\bm p_i|$ and
$\cos\theta_{12}\equiv\widehat{\bm p}_1\cdot\widehat{\bm p}_2\in[-1,1]$
is the opening-angle cosine. The electron-energy integrals cover the
kinematically allowed continuum, $\varepsilon_i\geq m_e$, subject to the
delta function in Eq.~\eqref{phase-space-measure}; nuclear recoil in the
energy balance is neglected. The differential rate takes the form
\begin{align}
\dd\Gamma_{0\nu}
=\frac{a_{0\nu}^{\rm rate}}{(m_eR)^2}
\left[A_0^{SP}-B_0^{SP}
(\widehat{\bm p}_1\cdot\widehat{\bm p}_2)\right]\dd\Omega_{0\nu},
\label{decay-rate-SP}
\end{align}
where $A_0^{SP}$ and $B_0^{SP}$ are the isotropic and angular-correlation
functions, respectively.  The superscript $SP$ indicates that the scalar
terms and their interference with the standard mass mechanism are included.
The rate coefficient $a_{0\nu}^{\rm rate}$ is distinct from the amplitude
coefficient $C_{0\nu}$ in Eq.~\eqref{C0nu-amplitude}.

The intermediate electron-amplitude coefficients are
\begin{align}
A_0^{SP}&=\sum_{i=1}^{4}|M_i|^2,
\label{A-SP}\\
B_0^{SP}&=\operatorname{Re}\left(
M_1M_2^*+M_1^*M_2+M_3M_4^*+M_3^*M_4\right),
\label{B-SP}\\
M_1&=\alpha_{-1-1}^{*}\Bigg\{
\left[Z_1^X+\frac{2}{m_eR}C_{4R}^{B}\right]
+\frac16\left(\frac{\zeta}{m_eR}-2\right)C_2^A\Bigg\},
\label{M1-SP}\\
M_2&=\alpha_{+1+1}^{*}\Bigg\{
\left[Z_1^X+\frac{2}{m_eR}C_{4R}^{B}\right]
+\frac16\left(\frac{\zeta}{m_eR}+2\right)C_2^A\Bigg\},
\label{M2-SP}\\
M_3&=\alpha_{+1-1}^{*}\Bigg\{
\left[Z_1^X+\frac{2}{m_eR}C_{4R}^{B}\right]
+\frac16\frac{\zeta}{m_eR}C_2^A\Bigg\},
\label{M3-SP}\\
M_4&=\alpha_{-1+1}^{*}\Bigg\{
\left[Z_1^X+\frac{2}{m_eR}C_{4R}^{B}\right]
+\frac16\frac{\zeta}{m_eR}C_2^A\Bigg\}.
\label{M4-SP}
\end{align}
The square-bracketed term in each $M_i$ contains the
$s_{1/2}$--$s_{1/2}$ contribution, while the term proportional to $C_2^A$
contains the $s_{1/2}$--$p_{1/2}$ contribution. The Coulomb factors are
\begin{align}
\alpha_{ij}&\equiv\widetilde A_i(\varepsilon_2)
\widetilde A_j(\varepsilon_1),
&
\zeta&\equiv3\alpha Z_f+(\varepsilon_1+\varepsilon_2)R,
\label{alpha-zeta}
\end{align}
where $i,j\in\{-1,+1\}$, $\widetilde A_{\pm1}$ is defined in
Appendix~\ref{elec-wavefunc},
$Z_i$ and $Z_f=Z_i+2$ are the atomic numbers of the parent and daughter
nuclei, respectively, and $\alpha$ is the
fine-structure constant (distinct from the indexed products $\alpha_{ij}$).
Equations~\eqref{M1-SP}--\eqref{M4-SP} already implement the dominant
scalar-recoil approximation specified in Sec.~\ref{NME}; the subleading
$C_{4R}^{A}$ contribution is therefore absent.

\subsection{Integrated half-life}
The two scalar couplings and the mass parameter may be complex.  Their
relative phases are
\begin{align}
\psi_3&=\operatorname{arg}\!\left[\eta_\nu(\epsilon_+)^*\right],
&
\psi_4&=\operatorname{arg}\!\left[\eta_\nu(\epsilon_-)^*\right],
\nonumber\\
\psi_{+-}&=\operatorname{arg}\!\left[\epsilon_-(\epsilon_+)^*\right]
=\psi_3-\psi_4\pmod{2\pi}.
\label{phases-SP}
\end{align}
After integration over the electron phase space, the inverse half-life,
$[T_{1/2}^{0\nu}]^{-1}\equiv\Gamma_{0\nu}/\ln2$, is
\begin{align}
\left[T_{1/2}^{0\nu}\right]^{-1}
=|M_{GT}^{0\nu}|^2\Big[&
|\eta_\nu|^2C_1
+4|\eta_\nu||\epsilon_-|C_2^{SP}\cos\psi_4
+4|\eta_\nu||\epsilon_+|C_{2+}^{SP}\cos\psi_3
\nonumber\\
&+4|\epsilon_-|^2C_3^{SP}
+4|\epsilon_+|^2C_{3+}^{SP}
+8|\epsilon_-||\epsilon_+|C_{+-}^{SP}\cos\psi_{+-}\Big].
\label{half-life-SP}
\end{align}
In the conventional real-NME phase choice,
\begin{align}
C_1&=(\chi_F-1)^2G_{01},
\label{C1}\\
C_2^{SP}
&=-(\chi_F-1)(\chi_B^{\prime SP}+\chi_D^{\prime SP})G_{05}^{SP}
+(\chi_F-1)\chi_F^{\prime SP}G_{06}^{SP},
\label{C2SP}\\
C_{2+}^{SP}
&=(\chi_F-1)(\chi_B^{\prime SP}-\chi_D^{\prime SP})G_{05}^{SP}
+(\chi_F-1)\chi_F^{\prime SP}G_{06}^{SP},
\label{C2+SP}\\
C_3^{SP}
&=(\chi_B^{\prime SP}+\chi_D^{\prime SP})^2G_{02}^{SP}
-(\chi_B^{\prime SP}+\chi_D^{\prime SP})
\chi_F^{\prime SP}G_{03}^{SP}
+(\chi_F^{\prime SP})^2G_{04}^{SP},
\label{C3SP}\\
C_{3+}^{SP}
&=(\chi_B^{\prime SP}-\chi_D^{\prime SP})^2G_{02}^{SP}
+(\chi_B^{\prime SP}-\chi_D^{\prime SP})
\chi_F^{\prime SP}G_{03}^{SP}
+(\chi_F^{\prime SP})^2G_{04}^{SP},
\label{C3+SP}\\
C_{+-}^{SP}
&=-\left[(\chi_B^{\prime SP})^2-(\chi_D^{\prime SP})^2\right]
G_{02}^{SP}
-\chi_F^{\prime SP}\chi_D^{\prime SP}G_{03}^{SP}
+(\chi_F^{\prime SP})^2G_{04}^{SP}.
\label{C+-SP}
\end{align}
All $C_i$ in Eq.~\eqref{half-life-SP} have dimensions of inverse time.

The three scalar NME ratios are
\begin{align}
\chi_F^{\prime SP}
&=\frac{F_S}{g_V}\left(\frac{g_V}{g_A}\right)^2
\frac{M_F'}{M_{GT}^{0\nu}},
\label{chi-F-SP}\\
\chi_B^{\prime SP}
&=\frac{F_{PS}}{g_A}\frac{M_B'}{M_{GT}^{0\nu}},
\label{chi-B-SP}\\
\chi_D^{\prime SP}
&=\frac{F_S}{g_A}\frac{g_V}{g_A}
\frac{M_D'}{M_{GT}^{0\nu}},
\label{chi-D-SP}
\end{align}
Here $F_S\equiv F_S(0)=g_S$ and $F_{PS}\equiv F_{PS}(0)=g_{PS}$ are the
zero-momentum reference values factored from the scalar NME ratios. If their
full $Q^2$ dependence is retained, it is included
inside the corresponding neutrino-potential integrals rather than also
being pulled outside the matrix elements. Equivalently, the kernels defining
$M_F'$ and $M_D'$ are weighted by $F_S(Q^2)/F_S(0)$, while that defining
$M_B'$ is weighted by $F_{PS}(Q^2)/F_{PS}(0)$.
The corresponding matrix elements are
\begin{align}
M_F'&=\dblmatel{0_f^+}{
\sum_{n\ne m}\tau_n^+\tau_m^+h_+'(r_{nm})}{0_i^+},
\nonumber\\
M_B'&=\dblmatel{0_f^+}{
\sum_{n\ne m}\tau_n^+\tau_m^+
\frac{iR}{2r_{nm}}h_+'(r_{nm})
\widehat{\bm r}_{nm}\cdot
(\bm\sigma_nB_m-\bm\sigma_mB_n)}{0_i^+},
\nonumber\\
M_D'&=\dblmatel{0_f^+}{
\sum_{n\ne m}\tau_n^+\tau_m^+
\frac{iR}{2r_{nm}}h_+'(r_{nm})
\widehat{\bm r}_{nm}\cdot(\bm D_n-\bm D_m)}{0_i^+}.
\label{scalar-NMEs}
\end{align}
The labels $F$, $B$, and $D$ on $M_F'$, $M_B'$, and $M_D'$ identify,
respectively, the Fermi-type operator, the pseudoscalar recoil operator
$B_n$, and the vector recoil operator $\bm D_n$. The label $F$ in
$\chi_F^{\prime SP}$ emphasizes its Fermi-type nuclear
operator.  The same quantity is denoted $\chi_P^{\prime SP}$ in
Refs.~\cite{PhysRevD.76.093009,PhysRevD.105.099902}, where $P$ labels the
electron $P$ wave.  The surviving operator combinations become
\begin{align}
C_2^A(\epsilon_\pm)
&=2M_{GT}^{0\nu}\epsilon_\pm\chi_F^{\prime SP},
\nonumber\\
C_{4R}^B(\epsilon_-)
&=-M_{GT}^{0\nu}\epsilon_-
(\chi_B^{\prime SP}+\chi_D^{\prime SP}),
\nonumber\\
C_{4R}^B(\epsilon_+)
&=M_{GT}^{0\nu}\epsilon_+
(\chi_B^{\prime SP}-\chi_D^{\prime SP}).
\label{operator-to-NME}
\end{align}
Here $C_2^A(\epsilon_\pm)$ and $C_{4R}^B(\epsilon_\pm)$ denote the parts
of the corresponding operator coefficients that are linear in the stated
scalar coupling.

\subsection{Simultaneous scalar interactions}
To expose the scalar--scalar interference directly, define
\begin{align}
\mathcal B_-&\equiv\chi_B^{\prime SP}+\chi_D^{\prime SP},
&
\mathcal B_+&\equiv\chi_B^{\prime SP}-\chi_D^{\prime SP},
&
\mathcal F&\equiv\chi_F^{\prime SP},
\end{align}
and
\begin{align}
X_B&=-\mathcal B_-\epsilon_-+\mathcal B_+\epsilon_+,
&
X_F&=\mathcal F(\epsilon_-+\epsilon_+).
\label{XB-XF}
\end{align}
Then
\begin{align}
C_{4R}^B&=M_{GT}^{0\nu}X_B,
&
C_2^A&=2M_{GT}^{0\nu}X_F,
\end{align}
and the half-life can be written as
\begin{align}
\left[T_{1/2}^{0\nu}\right]^{-1}
=|M_{GT}^{0\nu}|^2\Bigg\{&
|\eta_\nu|^2C_1
+4(\chi_F-1)\operatorname{Re}\!\left[
\eta_\nu\left(X_B^*G_{05}^{SP}+X_F^*G_{06}^{SP}\right)\right]
\nonumber\\
&+4\left[
|X_B|^2G_{02}^{SP}
+\operatorname{Re}(X_BX_F^*)G_{03}^{SP}
+|X_F|^2G_{04}^{SP}\right]\Bigg\}.
\label{half-life-coherent}
\end{align}
The pure-scalar quadratic form satisfies
\begin{align}
&|X_B|^2G_{02}^{SP}
+\operatorname{Re}(X_BX_F^*)G_{03}^{SP}
+|X_F|^2G_{04}^{SP}
\nonumber\\
&\qquad=
|\epsilon_-|^2C_3^{SP}
+|\epsilon_+|^2C_{3+}^{SP}
+2\operatorname{Re}(\epsilon_-\epsilon_+^*)C_{+-}^{SP}.
\label{scalar-polarization}
\end{align}
Expanding Eq.~\eqref{scalar-polarization} gives Eq.~\eqref{C+-SP} and fixes
the factor of eight in Eq.~\eqref{half-life-SP}.

The isotropic phase-space factors are
\begin{align}
G_{0k}^{SP}
&=\frac{a_{0\nu}^{\rm rate}}{\ln2\,(m_eR)^2}
\int a_{0k}^{SP}\,\dd\Omega_{0\nu},
\qquad k=2,\ldots,6,
\nonumber\\
G_{01}
&=\frac{a_{0\nu}^{\rm rate}}{\ln2\,(m_eR)^2}
\int a_{01}\,\dd\Omega_{0\nu},
\qquad a_{01}=\alpha_++\beta_+.
\label{PSFG_ok-SP}
\end{align}
The functions $a_{0k}^{SP}$ and the angular functions $b_{0k}^{SP}$ are
given in Appendix~\ref{scalar-spin-sums}.  The terms proportional to
$b_{0k}^{SP}$ vanish
after integration over
$\widehat{\bm p}_1\cdot\widehat{\bm p}_2$.

\subsection{Consistency checks}
The two single-coupling limits are
\begin{align}
\left.\left[T_{1/2}^{0\nu}\right]^{-1}\right|_{\epsilon_+=0}
&=|M_{GT}^{0\nu}|^2\left[
|\eta_\nu|^2C_1
+4|\eta_\nu||\epsilon_-|C_2^{SP}\cos\psi_4
+4|\epsilon_-|^2C_3^{SP}\right],
\label{minus-limit}\\
\left.\left[T_{1/2}^{0\nu}\right]^{-1}\right|_{\epsilon_-=0}
&=|M_{GT}^{0\nu}|^2\left[
|\eta_\nu|^2C_1
+4|\eta_\nu||\epsilon_+|C_{2+}^{SP}\cos\psi_3
+4|\epsilon_+|^2C_{3+}^{SP}\right].
\label{plus-limit}
\end{align}
These reproduce the corrected single-coupling scalar $P$-wave expressions
of Refs.~\cite{PhysRevD.76.093009,PhysRevD.105.099902}.

For the pure-scalar rate, define
\begin{align}
\mathsf G_{SP}
&\equiv
\begin{pmatrix}
G_{02}^{SP}&G_{03}^{SP}/2\\
G_{03}^{SP}/2&G_{04}^{SP}
\end{pmatrix},
&
\bm v_-&\equiv
\begin{pmatrix}-\mathcal B_-\\ \mathcal F\end{pmatrix},
&
\bm v_+&\equiv
\begin{pmatrix}\mathcal B_+\\ \mathcal F\end{pmatrix}.
\end{align}
In the real-NME convention used above, $\bm v_\pm$ are real two-component
vectors and the superscript $T$ below denotes ordinary transpose.
Positivity of the spin-summed rate requires
\begin{align}
G_{02}^{SP}&\geq0,
&
G_{04}^{SP}&\geq0,
&
4G_{02}^{SP}G_{04}^{SP}-(G_{03}^{SP})^2&\geq0.
\label{PSF-positivity}
\end{align}
Moreover,
\begin{align}
C_3^{SP}&=\bm v_-^T\mathsf G_{SP}\bm v_-,
&
C_{3+}^{SP}&=\bm v_+^T\mathsf G_{SP}\bm v_+,
&
C_{+-}^{SP}&=\bm v_-^T\mathsf G_{SP}\bm v_+,
\end{align}
so the Cauchy--Schwarz inequality gives
\begin{align}
|C_{+-}^{SP}|^2\leq C_3^{SP}C_{3+}^{SP}.
\label{interference-bound}
\end{align}

This bound is a special case, restricted to the two long-range scalar
couplings and derived here in closed analytic form, of the broader
question of which combinations of $0\nu\beta\beta$ mechanisms are
distinguishable from measured decay rates---a question addressed more
generally, across all 32 low-energy effective operators and via
cross-isotope comparison, in Ref.~\cite{GrafLindnerScholer2022}, where
the limiting factor is identified as the currently uncertain
low-energy constants rather than the rate formula itself. A numerical
implementation of the general EFT framework is provided by the
$\nu$DoBe package \cite{ScholerDeVriesGraf2023}.

\section{Summary \& Outlook}
We have derived the long-range amplitudes generated by a standard $V-A$
vertex and the two scalar $S\mp P$ hadronic currents coupled to a common
$S+P$ leptonic current.  Nuclear recoil and the
$s_{1/2}$--$p_{1/2}$ electron partial waves generate the three scalar NME
ratios $\chi_F^{\prime SP}$, $\chi_B^{\prime SP}$, and
$\chi_D^{\prime SP}$.  The first contains a Fermi-type operator with the
potential $h_+'$, while the other two contain the pseudoscalar operator
$B_n$ and recoil operator $\bm D_n$, respectively.

Both scalar interactions have been retained coherently.  Their mutual
interference contributes
$8|\epsilon_-||\epsilon_+|C_{+-}^{SP}\cos\psi_{+-}$ to the inverse
half-life, in addition to their separate interference with the standard mass
mechanism.  The single-coupling limits provide a cross-check of the corrected
scalar $P$-wave results
\cite{PhysRevD.76.093009,PhysRevD.105.099902}, and the bound in
Eq.~\eqref{interference-bound} supplies an internal positivity check.

The calculation uses light-neutrino exchange, closure, the single-nucleon
impulse approximation, the dominant scalar-recoil subset, and the
$s_{1/2}$ and $p_{1/2}$ Coulomb waves of the outgoing electrons.  The
explicit simultaneous-coupling formula provides a basis for correlated
constraints on the two scalar coefficients.  A numerical implementation
should retain the momentum dependence of the nucleon form factors and can
be extended to the pion-exchange and short-distance operators that arise in
chiral EFT \cite{Cirigliano:2018yza}.

\section*{Acknowledgments}
The author gratefully acknowledges the Physics Department at Central Michigan University for access to its research resources.

\appendix

\section{Calculation of the scalar amplitude}\label{scalarterms}
This appendix gives the operator reduction leading from the scalar
$\mathcal S$-matrix element in Eq.~\eqref{Ss-ps+p} to the closure-separated
amplitudes in Eqs.~\eqref{MqSPn} and \eqref{MqSPc}.  We use
\begin{align}
\Psi_1(\bm x_\alpha)&\equiv
\Psi_{s_1}(\varepsilon_1,\bm x_\alpha),
&\Psi_2(\bm x_\alpha)&\equiv
\Psi_{s_2}(\varepsilon_2,\bm x_\alpha),
\nonumber\\
\Delta_\alpha&\equiv E_a-E_i+\varepsilon_\alpha,
&\alpha&=1,2,
\nonumber\\
s^{\rho\mu}(1\bm x_1,2\bm x_2)
&\equiv\overline\Psi_1(\bm x_1)
\gamma^\rho\gamma^\mu(1+\gamma_5)\Psi_2^C(\bm x_2),
\nonumber\\
s^{\rho\mu}(2\bm x_1,1\bm x_2)
&\equiv\overline\Psi_2(\bm x_1)
\gamma^\rho\gamma^\mu(1+\gamma_5)\Psi_1^C(\bm x_2).
\label{app-s-rhomu}
\end{align}
The combined scalar current is the one in
Eq.~\eqref{combined-scalar-current}.  In this notation,
\begin{align}
\matrixel{f}{\mathcal S_{0\nu}^{(2)}}{i}^{SP}
={}&G_\beta^2\int\dd[3]{\bm x_1}\dd[3]{\bm x_2}
\int\frac{\dd[3]{\bm q}}{(2\pi)^3}
\frac{e^{i\bm q\cdot(\bm x_1-\bm x_2)}}{2\omega}
\sum_a2\pi i\delta(E_i-E_f-\varepsilon_1-\varepsilon_2)\,q_\mu
\nonumber\\
&\times\Bigg\{
\left[
\frac{s^{\rho\mu}(1\bm x_1,2\bm x_2)}{\omega+\Delta_2}
-\frac{s^{\rho\mu}(2\bm x_1,1\bm x_2)}{\omega+\Delta_1}
\right]
\left\langle J_{V-A,\rho}^\dagger(\bm x_1)
\widetilde J_{SP}^\dagger(\bm x_2)\right\rangle
\nonumber\\
&\qquad-
\left[
\frac{s^{\mu\rho}(1\bm x_1,2\bm x_2)}{\omega+\Delta_2}
-\frac{s^{\mu\rho}(2\bm x_1,1\bm x_2)}{\omega+\Delta_1}
\right]
\left\langle\widetilde J_{SP}^\dagger(\bm x_1)
J_{V-A,\rho}^\dagger(\bm x_2)\right\rangle
\Bigg\}.
\label{S-scalar}
\end{align}
The brackets in Eq.~\eqref{S-scalar} denote products of nuclear transition
matrix elements through an intermediate state:
\begin{align}
\left\langle J_A^\dagger(\bm x_1)J_B^\dagger(\bm x_2)\right\rangle
\equiv
\matrixel{N_f}{J_A^\dagger(\bm x_1)}{N_a}
\matrixel{N_a}{J_B^\dagger(\bm x_2)}{N_i}.
\end{align}

For a factorized two-nucleon operator, define the transition-density
functional
\begin{align}
T_a[\mathcal O_n\mathcal O_m]\equiv g_A^2\sum_{n\ne m}
&\matrixel{N_f}{
\tau_n^+\delta^{(3)}(\bm x_1-\bm r_n)\mathcal O_n}{N_a}
\nonumber\\
\times{}&
\matrixel{N_a}{
\tau_m^+\delta^{(3)}(\bm x_2-\bm r_m)\mathcal O_m}{N_i}.
\label{app-Ta}
\end{align}
It acts linearly on sums of factorized pair operators.  Using the
nonrelativistic currents in Eqs.~\eqref{Js+/-pnr} and \eqref{Jv-anr}, the
two current orderings become
\begin{align}
\left\langle J_{V-A,\rho}^\dagger(\bm x_1)
\widetilde J_{SP}^\dagger(\bm x_2)\right\rangle
=T_a\Big[&
g_{\rho0}\big(
G_V^0\varepsilon_S'\mathbbm1_n\mathbbm1_m
+G_V^0\varepsilon_P'\mathbbm1_nB_m
-\varepsilon_S'C_n\mathbbm1_m\big)
\nonumber\\
&+g_{\rho l}\big(
\varepsilon_S'\sigma_n^l\mathbbm1_m
+\varepsilon_P'\sigma_n^lB_m
-G_V^0\varepsilon_S'D_n^l\mathbbm1_m\big)\Big],
\label{app-hadronic-order-1}\\
\left\langle\widetilde J_{SP}^\dagger(\bm x_1)
J_{V-A,\rho}^\dagger(\bm x_2)\right\rangle
=T_a\Big[&
g_{\rho0}\big(
G_V^0\varepsilon_S'\mathbbm1_n\mathbbm1_m
+G_V^0\varepsilon_P'B_n\mathbbm1_m
-\varepsilon_S'\mathbbm1_nC_m\big)
\nonumber\\
&+g_{\rho l}\big(
\varepsilon_S'\mathbbm1_n\sigma_m^l
+\varepsilon_P'B_n\sigma_m^l
-G_V^0\varepsilon_S'\mathbbm1_nD_m^l\big)\Big].
\label{app-hadronic-order-2}
\end{align}
Here $G_V^0$, $\varepsilon_S'$, and $\varepsilon_P'$ are defined in
Eq.~\eqref{SP-comp-coupl}.  Terms containing two recoil operators,
$B_nC_m$ or $B_n\bm D_m$, are omitted.

Separating $q^\mu=(\omega,\bm q)$ gives
\begin{align}
\matrixel{f}{\mathcal S_{0\nu}^{(2)}}{i}^{SP}
={}&\frac{G_\beta^2}{8\pi}
\int\dd[3]{\bm x_1}\dd[3]{\bm x_2}
\int\frac{\dd[3]{\bm q}}{2\pi^2}
\frac{e^{i\bm q\cdot(\bm x_1-\bm x_2)}}{\omega}
\sum_a2\pi i\delta(E_i-E_f-\varepsilon_1-\varepsilon_2)
\nonumber\\
&\times\left[\omega\,\mathcal K_\omega-q^k\mathcal K_k\right],
\label{S-scalar-split}
\end{align}
where
\begin{align}
\mathcal K_\omega={}&
\left[
\frac{s^{\rho0}(1\bm x_1,2\bm x_2)}{\omega+\Delta_2}
-\frac{s^{\rho0}(2\bm x_1,1\bm x_2)}{\omega+\Delta_1}
\right]
\left\langle J_{V-A,\rho}^\dagger\widetilde J_{SP}^\dagger\right\rangle
\nonumber\\
&-
\left[
\frac{s^{0\rho}(1\bm x_1,2\bm x_2)}{\omega+\Delta_2}
-\frac{s^{0\rho}(2\bm x_1,1\bm x_2)}{\omega+\Delta_1}
\right]
\left\langle\widetilde J_{SP}^\dagger J_{V-A,\rho}^\dagger\right\rangle,
\nonumber\\
\mathcal K_k={}&
\left[
\frac{s^{\rho k}(1\bm x_1,2\bm x_2)}{\omega+\Delta_2}
-\frac{s^{\rho k}(2\bm x_1,1\bm x_2)}{\omega+\Delta_1}
\right]
\left\langle J_{V-A,\rho}^\dagger\widetilde J_{SP}^\dagger\right\rangle
\nonumber\\
&-
\left[
\frac{s^{k\rho}(1\bm x_1,2\bm x_2)}{\omega+\Delta_2}
-\frac{s^{k\rho}(2\bm x_1,1\bm x_2)}{\omega+\Delta_1}
\right]
\left\langle\widetilde J_{SP}^\dagger J_{V-A,\rho}^\dagger\right\rangle.
\end{align}
The spatial arguments of the hadronic currents in the last equations are
the same as in Eq.~\eqref{S-scalar}.

The needed gamma-matrix contractions are
\begin{align}
\gamma^\rho\gamma^0g_{\rho0}
&=\gamma^0\gamma^\rho g_{\rho0}=1,
&
\gamma^\rho\gamma^0g_{\rho l}
&=-\gamma^0\gamma^\rho g_{\rho l}
=-\gamma^l\gamma^0,
\nonumber\\
\gamma^\rho\gamma^kg_{\rho0}
&=\gamma^0\gamma^k,
&
\gamma^k\gamma^\rho g_{\rho0}
&=-\gamma^0\gamma^k,
\nonumber\\
\gamma^\rho\gamma^kg_{\rho l}
&=-\gamma^l\gamma^k,
&
\gamma^k\gamma^\rho g_{\rho l}
&=-\gamma^k\gamma^l.
\label{app-gamma-contractions}
\end{align}
For the electron bilinears, let
\begin{align}
u(12)&=\overline\Psi_1(\bm x_1)\Psi_2^C(\bm x_2),
&
u(21)&=\overline\Psi_1(\bm x_2)\Psi_2^C(\bm x_1),
\nonumber\\
u_5(12)&=\overline\Psi_1(\bm x_1)\gamma_5\Psi_2^C(\bm x_2),
&
u_5(21)&=\overline\Psi_1(\bm x_2)\gamma_5\Psi_2^C(\bm x_1),
\nonumber\\
u^{l0}(12)&=\overline\Psi_1(\bm x_1)\gamma^l\gamma^0\Psi_2^C(\bm x_2),
&
u^{l0}(21)&=\overline\Psi_1(\bm x_2)\gamma^l\gamma^0\Psi_2^C(\bm x_1),
\nonumber\\
u_5^{l0}(12)&=\overline\Psi_1(\bm x_1)
\gamma^l\gamma^0\gamma_5\Psi_2^C(\bm x_2),
&
u_5^{l0}(21)&=\overline\Psi_1(\bm x_2)
\gamma^l\gamma^0\gamma_5\Psi_2^C(\bm x_1),
\label{app-u-bilinears}\\
F_\pm&=\frac12[u(12)\pm u(21)],
&
F_{5\pm}&=\frac12[u_5(12)\pm u_5(21)],
\nonumber\\
F_\pm^{l0}&=\frac12[u^{l0}(12)\pm u^{l0}(21)],
&
F_{5\pm}^{l0}&=\frac12[u_5^{l0}(12)\pm u_5^{l0}(21)],
\nonumber\\
E_\pm&=F_\pm+F_{5\pm},
&
E_\pm^l&=F_\pm^{l0}+F_{5\pm}^{l0}.
\label{app-electron-currents}
\end{align}
These definitions reproduce Eqs.~\eqref{Epm-definition} and
\eqref{Epml-definition}. The local symbols $F_\pm$ and $F_{5\pm}$ in this
appendix denote electron bilinears and are unrelated to the nucleon form
factors $F_S$ and $F_{PS}$.

Using the neutrino potentials in Eqs.~\eqref{neu-pot-energy} and
\eqref{neu-pot-mom}, the energy-numerator contribution reduces to
\begin{align}
\mathcal I_\omega=T_a\Big\{&
(H_{\omega2}-H_{\omega1})
\left[-(A_4^l+A_{4R}^l)E_+^l+B_{2R}E_-\right]
\nonumber\\
&+(H_{\omega2}+H_{\omega1})
\left[-(A_4^l+A_{4R}^l)E_-^l+B_{2R}E_+\right]\Big\},
\label{energy-term-SP}
\end{align}
and the three-momentum contribution becomes
\begin{align}
\mathcal I_q=T_a\Big\{&
(H_{q2}^l+H_{q1}^l)
\left[-(A_2+A_{2R})E_-^l
+(A_5^{lk}+A_{5R}^{lk})E_-^k
-(B_4^l+B_{4R}^l)E_+\right]
\nonumber\\
&+(H_{q2}^l-H_{q1}^l)
\left[-(A_2+A_{2R})E_+^l
+(A_5^{lk}+A_{5R}^{lk})E_+^k
-(B_4^l+B_{4R}^l)E_-\right]\Big\}.
\label{mom-term-SP}
\end{align}
Apart from the common integrations and normalization in
Eq.~\eqref{S-scalar-split}, $\mathcal I_\omega$ and $\mathcal I_q$ are the
contributions generated by $\omega\mathcal K_\omega$ and
$-q^k\mathcal K_k$, respectively.
The nuclear combinations in these equations are defined in
Eqs.~\eqref{pair-operators} and \eqref{A_B-SP}.  Sorting
Eqs.~\eqref{energy-term-SP} and \eqref{mom-term-SP} by exchange parity gives
the nonvanishing and closure-odd pieces displayed in
Eqs.~\eqref{MqSPn} and \eqref{MqSPc}.

\section{Electron wavefunctions}\label{elec-wavefunc}
The emitted electrons are described by solutions of the Dirac equation in
the Coulomb field of the daughter nucleus.  A continuum state with energy
$\varepsilon$, asymptotic momentum direction $\widehat{\bm p}$, and spin
projection $s$ can be expanded in spherical waves
\cite{rose1961relativistic,Doi+Kotani1985}:
\begin{align}
\Psi_s(\varepsilon,\bm r,\widehat{\bm p})
&=\sum_{\kappa,m}
a_{\kappa m}(\widehat{\bm p},s)
\psi_{\kappa m}(\varepsilon,\bm r)
\nonumber\\
&=\Psi_s^{s_{1/2}}+\Psi_s^{p_{1/2}}
+\Psi_s^{p_{3/2}}+\cdots .
\label{Psi}
\end{align}
Here $\kappa$ is the Dirac angular quantum number,
$j_\kappa=|\kappa|-\tfrac12$, and $m$ is the magnetic projection of
$j_\kappa$. The corresponding orbital angular momentum is
$\ell_\kappa=\kappa$ for $\kappa>0$ and
$\ell_\kappa=-\kappa-1$ for $\kappa<0$. With this notation,
\begin{align}
a_{\kappa m}(\widehat{\bm p},s)
=4\pi i^{\ell_\kappa}
C\!\left(\ell_\kappa,\frac12,j_\kappa
\middle|m-s,s\right)
\left[Y_{\ell_\kappa}^{m-s}(\widehat{\bm p})\right]^*,
\label{weight-fact-a}
\end{align}
where $C(\,|\,)$ is a Clebsch--Gordan coefficient and $Y_\ell^m$ is a
spherical harmonic. The stationary
spherical spinor is
\begin{align}
\psi_{\kappa m}(\varepsilon,\bm r)
=
\begin{pmatrix}
g_\kappa(\varepsilon,r)\,
\chi_{\kappa m}(\widehat{\bm r})\\
i f_\kappa(\varepsilon,r)\,
\chi_{-\kappa m}(\widehat{\bm r})
\end{pmatrix},
\label{DiracWaveFunc}
\end{align}
with
\begin{align}
\chi_{\kappa m}(\widehat{\bm r})
=\sum_{m_s=\pm1/2}
C\!\left(\ell_\kappa,\frac12,j_\kappa
\middle|m-m_s,m_s\right)
Y_{\ell_\kappa}^{m-m_s}(\widehat{\bm r})\chi_{m_s}.
\label{chi}
\end{align}
The two-component Pauli spinor $\chi_{m_s}$ has spin projection $m_s$;
$g_\kappa$ and $f_\kappa$ are the upper- and lower-component radial
functions, respectively.

Only the $s_{1/2}$ and $p_{1/2}$ waves are retained. The Dirac quantum
numbers are $\kappa=-1$ for $s_{1/2}$ and $\kappa=+1$ for $p_{1/2}$.
Their convenient four-component forms are
\begin{align}
\Psi_s^{s_{1/2}}(\varepsilon,\bm r,\widehat{\bm p})
&=
\begin{pmatrix}
\widetilde g_{-1}(\varepsilon,r)\chi_s\\
\widetilde f_{+1}(\varepsilon,r)
(\bm\sigma\cdot\widehat{\bm p})\chi_s
\end{pmatrix},
\label{Shalf}\\
\Psi_s^{p_{1/2}}(\varepsilon,\bm r,\widehat{\bm p})
&=i
\begin{pmatrix}
\widetilde g_{+1}(\varepsilon,r)
(\bm\sigma\cdot\widehat{\bm r})
(\bm\sigma\cdot\widehat{\bm p})\chi_s\\
-\widetilde f_{-1}(\varepsilon,r)
(\bm\sigma\cdot\widehat{\bm r})\chi_s
\end{pmatrix}.
\label{Phalf}
\end{align}
Here and below, $r=|\bm r|$,
$\widehat{\bm r}=\bm r/r$, and
$\widehat{\bm p}=\bm p/|\bm p|$.
To first order in $r/R$, the radial functions are approximated by
\begin{align}
\widetilde g_{-1}(\varepsilon,r)
&=\widetilde A_{-1}(\varepsilon),
&
\widetilde f_{+1}(\varepsilon,r)
&=\widetilde A_{+1}(\varepsilon),
\nonumber\\
\widetilde g_{+1}(\varepsilon,r)
&=\widetilde A_{-1}(\varepsilon)\xi_+(\varepsilon)\frac rR,
&
\widetilde f_{-1}(\varepsilon,r)
&=-\widetilde A_{+1}(\varepsilon)\xi_-(\varepsilon)\frac rR,
\label{g-f}\\
\widetilde A_{\pm1}(\varepsilon)
&=\sqrt{\frac{\varepsilon\mp m_e}{2\varepsilon}
F_0(Z_f,\varepsilon)},
&
\xi_\pm(\varepsilon)
&=\frac12\alpha Z_f+\frac13(\varepsilon\pm m_e)R.
\nonumber
\end{align}
Here $\varepsilon>m_e$ is the total electron energy; $Z_i$ and
$Z_f=Z_i+2$ are the parent- and daughter-nucleus atomic numbers,
respectively; $\alpha$ is the fine-structure constant; and
$R=r_0A^{1/3}$ is the nuclear radius used in the surface approximation.
The leading $s_{1/2}$ functions neglect
finite-de-Broglie-wavelength corrections. In the point-Coulomb
approximation evaluated at $r=R$, the relativistic Fermi function is
\begin{align}
F_0(Z_f,\varepsilon)
&=\frac{4}{\Gamma^2(2\gamma_1+1)}
(2pR)^{2(\gamma_1-1)}
\left|\Gamma(\gamma_1+iy)\right|^2e^{\pi y},
\nonumber\\
\gamma_1&=\sqrt{1-(\alpha Z_f)^2},
&
y&=\frac{\alpha Z_f\varepsilon}{p}.
\label{Fermi-function}
\end{align}
where
$p\equiv|\bm p|=\sqrt{\varepsilon^2-m_e^2}$ is the asymptotic electron
three-momentum magnitude, $\Gamma(z)$ is the Euler gamma function, and
$\gamma_1$ and $y$ are the dimensionless quantities defined in
Eq.~\eqref{Fermi-function}. This point-Coulomb expression neglects finite
nuclear-size and atomic-screening corrections
beyond the surface prescription \cite{Doi+Kotani1985,PhysRevD.76.093009}.

For the spin sums it is useful to write the Coulomb waves in terms of
free spinors. Here $p^\mu=(\varepsilon,\bm p)$ denotes the electron
four-momentum, whereas the unbold $p$ in Eq.~\eqref{Fermi-function} denotes
its spatial magnitude. We use unit-normalized spinors,
\begin{align}
u_s(p)&=\sqrt{\frac{\varepsilon+m_e}{2\varepsilon}}
\begin{pmatrix}
\chi_s\\[1mm]
\dfrac{\bm\sigma\cdot\bm p}{\varepsilon+m_e}\chi_s
\end{pmatrix},
&
v_s(p)&=\sqrt{\frac{\varepsilon+m_e}{2\varepsilon}}
\begin{pmatrix}
\dfrac{\bm\sigma\cdot\bm p}{\varepsilon+m_e}\chi_{-s}\\[1mm]
\chi_{-s}
\end{pmatrix}.
\label{u-v-spinors}
\end{align}
where $\chi_{-s}\equiv-i\sigma^2\chi_s^*$ is the charge-conjugate
two-component Pauli spinor (the overall phase is conventional). This
normalization gives
\begin{align}
\sum_su_s(p)\overline u_s(p)
&=\frac{\slashed p+m_e}{2\varepsilon},
&
\sum_sv_s(p)\overline v_s(p)
&=\frac{\slashed p-m_e}{2\varepsilon},
\label{spinor-completeness}
\end{align}
with $\slashed p\equiv\gamma^\mu p_\mu$.
Define the energy projectors
\begin{align}
P_{0\pm}\equiv\frac12(1\pm\gamma^0).
\label{P0pm}
\end{align}
The non-tilde radial functions are introduced through
\begin{align}
\widetilde g_{\pm1}(\varepsilon,r)
&=\sqrt{\frac{\varepsilon\mp m_e}{2\varepsilon}}\,
g_{\pm1}(\varepsilon,r),
\nonumber\\
\widetilde f_{\pm1}(\varepsilon,r)
&=\sqrt{\frac{\varepsilon\mp m_e}{2\varepsilon}}\,
f_{\pm1}(\varepsilon,r).
\label{tilde-nontilde-gf}
\end{align}
Equations~\eqref{Shalf} and \eqref{Phalf} can then be expressed as
\begin{align}
\Psi_s^{s_{1/2}}
&=\left[g_{-1}P_{0+}+f_{+1}P_{0-}\right]u_s(p),
\label{Shalf-u}\\
\Psi_s^{p_{1/2}}
&=-i\left[g_{+1}P_{0+}+f_{-1}P_{0-}\right]
\gamma_i\widehat r^iu_s(p).
\label{Phalf-u}
\end{align}
Their Dirac and charge conjugates are
\begin{align}
\overline{\Psi_s^{s_{1/2}}}
&=\overline u_s(p)\left[g_{-1}P_{0+}+f_{+1}P_{0-}\right],
\label{Shalf-ubar}\\
\left(\Psi_s^{s_{1/2}}\right)^C
&=\left[f_{+1}P_{0+}+g_{-1}P_{0-}\right]v_s(p),
\label{Shalf-uC}\\
\overline{\Psi_s^{p_{1/2}}}
&=-i\overline u_s(p)\gamma_i\widehat r^i
\left[g_{+1}P_{0+}+f_{-1}P_{0-}\right],
\label{Phalf-ubar}\\
\left(\Psi_s^{p_{1/2}}\right)^C
&=i\gamma_i\widehat r^i
\left[g_{+1}P_{0+}+f_{-1}P_{0-}\right]v_s(p).
\label{Phalf-uC}
\end{align}
The arguments $(\varepsilon,r)$ of the radial functions have been
suppressed in Eqs.~\eqref{Shalf-u}--\eqref{Phalf-uC}.

In the surface-factorization approximation,
\begin{align}
\widetilde g_{\pm1}(\varepsilon,R)
&\equiv\widetilde g_{\pm1}(\varepsilon),
&
\widetilde f_{\pm1}(\varepsilon,R)
&\equiv\widetilde f_{\pm1}(\varepsilon).
\end{align}
For two electrons we use
\begin{align}
\widetilde g_{\pm1}^{1}
&\equiv\widetilde g_{\pm1}(\varepsilon_1),
&
\widetilde g_{\pm1}^{2}
&\equiv\widetilde g_{\pm1}(\varepsilon_2),
\nonumber\\
\widetilde f_{\pm1}^{1}
&\equiv\widetilde f_{\pm1}(\varepsilon_1),
&
\widetilde f_{\pm1}^{2}
&\equiv\widetilde f_{\pm1}(\varepsilon_2),
\label{radial-shorthand}
\end{align}
with analogous notation for $g_{\pm1}$ and $f_{\pm1}$.

\section{Spin sums and phase-space factors}\label{PSF}
\subsection{General spin-sum identity}\label{Gen-Spinsum}
Consider two electron bilinears
\begin{align}
\Lambda_i&=\overline u_1\Gamma_i v_2,
&
\Lambda_j&=\overline u_1\Gamma_j v_2,
\end{align}
where $u_1=u_{s_1}(p_1)$ and $v_2=v_{s_2}(p_2)$.  For the currents needed
here,
\begin{align}
\Gamma_i&=\alpha P_{0+}+\beta P_{0-}
+\lambda\gamma_5P_{0+}+\eta\gamma_5P_{0-},
\nonumber\\
\Gamma_j&=\rho P_{0+}+\sigma P_{0-}
+\mu\gamma_5P_{0+}+\nu\gamma_5P_{0-},
\nonumber\\
\overline\Gamma_j
&=\gamma^0\Gamma_j^\dagger\gamma^0
=\rho P_{0+}+\sigma P_{0-}
-\mu P_{0+}\gamma_5-\nu P_{0-}\gamma_5.
\label{generic-Gamma}
\end{align}
The symbols $\alpha,\beta,\lambda,\eta,\rho,\sigma,\mu,$ and $\nu$ in
Eq.~\eqref{generic-Gamma} are generic real products of electron radial
functions in the phase convention used below; in this subsection only,
$\alpha$ therefore does not denote the fine-structure constant. The Dirac
adjoint of a matrix is
$\overline\Gamma_j\equiv\gamma^0\Gamma_j^\dagger\gamma^0$. Using the
spinor completeness relations in Eq.~\eqref{spinor-completeness}
\cite{griffiths2008introduction} gives
\begin{align}
\sum_{s_1,s_2}\Lambda_i\Lambda_j^*
=\frac{1}{4\varepsilon_1\varepsilon_2}
\operatorname{Tr}\!\left[
\Gamma_i(\slashed p_2-m_e)
\overline\Gamma_j(\slashed p_1+m_e)\right].
\label{Lambda-Lambda}
\end{align}
Expanding the trace,
\begin{align}
\operatorname{Tr}[
\Gamma_i\slashed p_2\overline\Gamma_j\slashed p_1]
={}&2\varepsilon_1\varepsilon_2
(\alpha\rho+\beta\sigma+\lambda\mu+\eta\nu)
\nonumber\\
&-2\bm p_1\cdot\bm p_2
(\alpha\sigma+\beta\rho+\lambda\nu+\eta\mu),
\nonumber\\
\operatorname{Tr}[
\Gamma_i\slashed p_2\overline\Gamma_j]
={}&2\varepsilon_2
(\alpha\rho-\beta\sigma-\lambda\mu+\eta\nu),
\nonumber\\
\operatorname{Tr}[
\Gamma_i\overline\Gamma_j\slashed p_1]
={}&2\varepsilon_1
(\alpha\rho-\beta\sigma+\lambda\mu-\eta\nu),
\nonumber\\
\operatorname{Tr}[
\Gamma_i\overline\Gamma_j]
={}&2(\alpha\rho+\beta\sigma-\lambda\mu-\eta\nu).
\label{spin-traces}
\end{align}
Substitution in Eq.~\eqref{Lambda-Lambda} yields
\begin{align}
\sum_{s_1,s_2}\Lambda_i\Lambda_j^*
=2\Big[&
\widetilde\alpha\widetilde\rho
+\widetilde\beta\widetilde\sigma
+\widetilde\lambda\widetilde\mu
+\widetilde\eta\widetilde\nu
\nonumber\\
&-\left(
\widetilde\alpha\widetilde\sigma
+\widetilde\beta\widetilde\rho
+\widetilde\lambda\widetilde\nu
+\widetilde\eta\widetilde\mu\right)
(\widehat{\bm p}_1\cdot\widehat{\bm p}_2)\Big].
\label{fund-spin}
\end{align}
For any
$x\in\{\alpha,\beta,\lambda,\eta,\rho,\sigma,\mu,\nu\}$, the notation
$\widetilde x$ means the same radial-function product as $x$, with every
$g_{\pm1}$ and $f_{\pm1}$ replaced by $\widetilde g_{\pm1}$ and
$\widetilde f_{\pm1}$ according to Eq.~\eqref{tilde-nontilde-gf}.

\subsection{Scalar electron currents and kinematic functions}
\label{scalar-spin-sums}
The three electron currents entering Eqs.~\eqref{MqnSP-ss} and
\eqref{MqnSP-sp} can be written as
\begin{align}
E_+^{s_{1/2}s_{1/2}}
&=\overline u_{s_1}(p_1)
\left[AP_{0+}+BP_{0-}
+D\gamma_5P_{0+}+C\gamma_5P_{0-}\right]v_{s_2}(p_2),
\nonumber\\
\widetilde E_+^{s_{1/2}p_{1/2}}
&=\overline u_{s_1}(p_1)
\Big[(E+F)P_{0+}-(G+H)P_{0-}
+(I-J)\gamma_5P_{0+}
+(K-L)\gamma_5P_{0-}\Big]v_{s_2}(p_2),
\nonumber\\
\widetilde E_-^{s_{1/2}p_{1/2}}
&=\overline u_{s_1}(p_1)
\Big[(E-F)P_{0+}+(G-H)P_{0-}
-(I+J)\gamma_5P_{0+}
+(K+L)\gamma_5P_{0-}\Big]v_{s_2}(p_2).
\label{scalar-spin-currents}
\end{align}
The capital Roman letters in Eq.~\eqref{scalar-spin-currents} denote
electron radial products:
\begin{align}
A&=g_{-1}^{1}f_{+1}^{2},
&B&=f_{+1}^{1}g_{-1}^{2},
&C&=g_{-1}^{1}g_{-1}^{2},
&D&=f_{+1}^{1}f_{+1}^{2},
\nonumber\\
E&=f_{-1}^{1}f_{+1}^{2},
&F&=g_{-1}^{1}g_{+1}^{2},
&G&=f_{+1}^{1}f_{-1}^{2},
&H&=g_{+1}^{1}g_{-1}^{2},
\nonumber\\
I&=f_{+1}^{1}g_{+1}^{2},
&J&=g_{+1}^{1}f_{+1}^{2},
&K&=f_{-1}^{1}g_{-1}^{2},
&L&=g_{-1}^{1}f_{-1}^{2}.
\label{ABCDEFGH}
\end{align}
The corresponding $\widetilde A,\ldots,\widetilde L$ are formed from the
tilded radial functions.
In the following expressions,
\begin{align}
\varepsilon_{21}&\equiv\varepsilon_2-\varepsilon_1=-\varepsilon_{12},
&
\zeta&\equiv3\alpha Z_f+(\varepsilon_1+\varepsilon_2)R,
\label{appendix-energy-combinations}
\end{align}
and the Coulomb products $\alpha_{ij}$ are defined in
Eq.~\eqref{alpha-zeta}.

Applying the general identity in Appendix~\ref{Gen-Spinsum} to the three
currents in Eq.~\eqref{scalar-spin-currents} gives the six spin sums
\begin{align}
\sum_{s_1,s_2}\left|E_+^{s_{1/2}s_{1/2}}\right|^2
&=2\left[C_{01}-D_{01}
(\widehat{\bm p}_1\cdot\widehat{\bm p}_2)\right],
\label{scalar-spin-sum-01}\\
\sum_{s_1,s_2}E_+^{s_{1/2}s_{1/2}}
\left(\widetilde E_+^{s_{1/2}p_{1/2}}\right)^*
&=2\left[C_{02}-D_{02}
(\widehat{\bm p}_1\cdot\widehat{\bm p}_2)\right],
\label{scalar-spin-sum-02}\\
\sum_{s_1,s_2}E_+^{s_{1/2}s_{1/2}}
\left(\widetilde E_-^{s_{1/2}p_{1/2}}\right)^*
&=2\left[C_{03}-D_{03}
(\widehat{\bm p}_1\cdot\widehat{\bm p}_2)\right],
\label{scalar-spin-sum-03}\\
\sum_{s_1,s_2}\widetilde E_+^{s_{1/2}p_{1/2}}
\left(\widetilde E_-^{s_{1/2}p_{1/2}}\right)^*
&=2\left[C_{04}-D_{04}
(\widehat{\bm p}_1\cdot\widehat{\bm p}_2)\right],
\label{scalar-spin-sum-04}\\
\sum_{s_1,s_2}\left|\widetilde E_+^{s_{1/2}p_{1/2}}\right|^2
&=2\left[C_{05}-D_{05}
(\widehat{\bm p}_1\cdot\widehat{\bm p}_2)\right],
\label{scalar-spin-sum-05}\\
\sum_{s_1,s_2}\left|\widetilde E_-^{s_{1/2}p_{1/2}}\right|^2
&=2\left[C_{06}-D_{06}
(\widehat{\bm p}_1\cdot\widehat{\bm p}_2)\right].
\label{scalar-spin-sum-06}
\end{align}
Using the radial products in Eq.~\eqref{ABCDEFGH}, the coefficients in
Eqs.~\eqref{scalar-spin-sum-01}--\eqref{scalar-spin-sum-06} are given
below. For each coefficient, the first form displays its dependence on
the radial products $\widetilde A,\ldots,\widetilde L$, while the second
form follows after inserting the surface radial functions in
Eq.~\eqref{g-f}:
\begin{align}
C_{01}
&=g_{01}
=\widetilde A^2+\widetilde B^2+\widetilde C^2+\widetilde D^2
\nonumber\\
&=|\alpha_{+1-1}|^2+|\alpha_{-1+1}|^2
+|\alpha_{-1-1}|^2+|\alpha_{+1+1}|^2,
\label{C01}\\
D_{01}
&=h_{01}=4\widetilde A\widetilde B
=4\alpha_{-1-1}\alpha_{+1+1}
=4\alpha_{+1-1}\alpha_{-1+1},
\label{D01}\\
C_{02}
&=\widetilde A\widetilde E+\widetilde A\widetilde F
-\widetilde B\widetilde G-\widetilde B\widetilde H
+\widetilde D\widetilde I-\widetilde D\widetilde J
+\widetilde C\widetilde K-\widetilde C\widetilde L
\nonumber\\
&=\frac{1}{3}(\zeta+\varepsilon_{21}R)
\left(\alpha_{+1-1}\alpha_{-1-1}
+\alpha_{-1+1}\alpha_{+1+1}\right)
\nonumber\\
&\quad-\frac{1}{3}(\zeta-\varepsilon_{21}R)
\left(\alpha_{+1-1}\alpha_{+1+1}
+\alpha_{-1+1}\alpha_{-1-1}\right),
\label{C02}\\
D_{02}
&=-\widetilde A\widetilde G-\widetilde A\widetilde H
+\widetilde B\widetilde E+\widetilde B\widetilde F
+\widetilde D\widetilde K-\widetilde D\widetilde L
+\widetilde C\widetilde I-\widetilde C\widetilde J
\nonumber\\
&=\frac{1}{3}\Big[
\alpha_{+1-1}\alpha_{+1+1}
(\zeta+\varepsilon_{21}R-2m_eR)
\nonumber\\
&\qquad
-\alpha_{+1-1}\alpha_{-1-1}
(\zeta-\varepsilon_{21}R+2m_eR)
\nonumber\\
&\qquad
-\alpha_{-1+1}\alpha_{+1+1}
(\zeta-\varepsilon_{21}R-2m_eR)
\nonumber\\
&\qquad
+\alpha_{-1+1}\alpha_{-1-1}
(\zeta+\varepsilon_{21}R+2m_eR)\Big],
\label{D02}\\
C_{03}
&=\widetilde A\widetilde E-\widetilde A\widetilde F
+\widetilde B\widetilde G-\widetilde B\widetilde H
-\widetilde D\widetilde I-\widetilde D\widetilde J
+\widetilde C\widetilde K+\widetilde C\widetilde L
\nonumber\\
&=-\frac{1}{3}(\zeta-\varepsilon_{21}R)
\left(\alpha_{+1-1}\alpha_{+1+1}
+\alpha_{-1+1}\alpha_{-1-1}\right)
\nonumber\\
&\quad-\frac{1}{3}(\zeta+\varepsilon_{21}R)
\left(\alpha_{+1-1}\alpha_{-1-1}
+\alpha_{-1+1}\alpha_{+1+1}\right),
\label{C03}\\
D_{03}
&=\widetilde A\widetilde G-\widetilde A\widetilde H
+\widetilde B\widetilde E-\widetilde B\widetilde F
+\widetilde D\widetilde K+\widetilde D\widetilde L
-\widetilde C\widetilde I-\widetilde C\widetilde J
\nonumber\\
&=-\frac{1}{3}\Big[
\alpha_{+1-1}\alpha_{+1+1}
(\zeta+\varepsilon_{21}R-2m_eR)
\nonumber\\
&\qquad
+\alpha_{+1-1}\alpha_{-1-1}
(\zeta-\varepsilon_{21}R+2m_eR)
\nonumber\\
&\qquad
+\alpha_{-1+1}\alpha_{+1+1}
(\zeta-\varepsilon_{21}R-2m_eR)
\nonumber\\
&\qquad
+\alpha_{-1+1}\alpha_{-1-1}
(\zeta+\varepsilon_{21}R+2m_eR)\Big],
\label{D03}\\
C_{04}
&=\widetilde E^2-\widetilde F^2-\widetilde G^2+\widetilde H^2
-\widetilde I^2+\widetilde J^2+\widetilde K^2-\widetilde L^2
\nonumber\\
&=\frac{1}{9}\Big\{
|\alpha_{+1-1}|^2\zeta(2m_eR-\varepsilon_{21}R)
\nonumber\\
&\qquad
-|\alpha_{-1+1}|^2\zeta(\varepsilon_{21}R+2m_eR)
\nonumber\\
&\qquad
-|\alpha_{-1-1}|^2\varepsilon_{21}R(\zeta+2m_eR)
\nonumber\\
&\qquad
-|\alpha_{+1+1}|^2\varepsilon_{21}R(\zeta-2m_eR)
\Big\},
\label{C04}\\
D_{04}
&=2\left(-\widetilde E\widetilde H+\widetilde F\widetilde G
+\widetilde I\widetilde L-\widetilde J\widetilde K\right)
\nonumber\\
&=-\frac{4}{9}\alpha_{+1-1}\alpha_{-1+1}
\zeta\,\varepsilon_{21}R,
\label{D04}\\
C_{05}
&=(\widetilde E+\widetilde F)^2+(\widetilde G+\widetilde H)^2
+(\widetilde I-\widetilde J)^2+(\widetilde K-\widetilde L)^2
\nonumber\\
&=\frac{1}{18}\Big\{
|\alpha_{+1+1}|^2
\left[(\zeta-2m_eR)^2+(\varepsilon_{21}R)^2\right]
\nonumber\\
&\qquad
+|\alpha_{-1-1}|^2
\left[(\zeta+2m_eR)^2+(\varepsilon_{21}R)^2\right]
\nonumber\\
&\qquad
+|\alpha_{-1+1}|^2
\left[\zeta^2+(\varepsilon_{21}R+2m_eR)^2\right]
\nonumber\\
&\qquad
+|\alpha_{+1-1}|^2
\left[\zeta^2+(\varepsilon_{21}R-2m_eR)^2\right]
\nonumber\\
&\qquad
-4\alpha_{+1-1}\alpha_{-1+1}
\left[\zeta^2-(\varepsilon_{21}R)^2\right]
\Big\},
\label{C05}\\
D_{05}
&=2\left[-(\widetilde E+\widetilde F)(\widetilde G+\widetilde H)
+(\widetilde I-\widetilde J)(\widetilde K-\widetilde L)\right]
\nonumber\\
&=\frac{1}{18}\Big\{
-|\alpha_{+1+1}|^2
\left[(\zeta-2m_eR)^2-(\varepsilon_{21}R)^2\right]
\nonumber\\
&\qquad
-|\alpha_{-1-1}|^2
\left[(\zeta+2m_eR)^2-(\varepsilon_{21}R)^2\right]
\nonumber\\
&\qquad
-|\alpha_{-1+1}|^2
\left[\zeta^2-(\varepsilon_{21}R+2m_eR)^2\right]
\nonumber\\
&\qquad
-|\alpha_{+1-1}|^2
\left[\zeta^2-(\varepsilon_{21}R-2m_eR)^2\right]
\nonumber\\
&\qquad
+4\alpha_{+1-1}\alpha_{-1+1}
\left[\zeta^2+(\varepsilon_{21}R)^2-(2m_eR)^2\right]
\Big\},
\label{D05}\\
C_{06}
&=(\widetilde E-\widetilde F)^2+(\widetilde G-\widetilde H)^2
+(\widetilde I+\widetilde J)^2+(\widetilde K+\widetilde L)^2
\nonumber\\
&=\frac{1}{18}\Big\{
|\alpha_{+1+1}|^2
\left[(\zeta-2m_eR)^2+(\varepsilon_{21}R)^2\right]
\nonumber\\
&\qquad
+|\alpha_{-1-1}|^2
\left[(\zeta+2m_eR)^2+(\varepsilon_{21}R)^2\right]
\nonumber\\
&\qquad
+|\alpha_{-1+1}|^2
\left[\zeta^2+(\varepsilon_{21}R+2m_eR)^2\right]
\nonumber\\
&\qquad
+|\alpha_{+1-1}|^2
\left[\zeta^2+(\varepsilon_{21}R-2m_eR)^2\right]
\nonumber\\
&\qquad
+4\alpha_{+1-1}\alpha_{-1+1}
\left[\zeta^2-(\varepsilon_{21}R)^2\right]
\Big\},
\label{C06}\\
D_{06}
&=2\Big[(\widetilde E-\widetilde F)(\widetilde G-\widetilde H)
-(\widetilde I+\widetilde J)(\widetilde K+\widetilde L)\Big]
\nonumber\\
&=\frac{1}{18}\Big\{
|\alpha_{+1+1}|^2
\left[(\zeta-2m_eR)^2-(\varepsilon_{21}R)^2\right]
\nonumber\\
&\qquad
+|\alpha_{-1-1}|^2
\left[(\zeta+2m_eR)^2-(\varepsilon_{21}R)^2\right]
\nonumber\\
&\qquad
+|\alpha_{-1+1}|^2
\left[\zeta^2-(\varepsilon_{21}R+2m_eR)^2\right]
\nonumber\\
&\qquad
+|\alpha_{+1-1}|^2
\left[\zeta^2-(\varepsilon_{21}R-2m_eR)^2\right]
\nonumber\\
&\qquad
+4\alpha_{+1-1}\alpha_{-1+1}
\left[\zeta^2+(\varepsilon_{21}R)^2-(2m_eR)^2\right]
\Big\}.
\label{D06}
\end{align}
The radial-product expressions follow term by term from
Eq.~\eqref{fund-spin}, and their explicit forms use only
Eqs.~\eqref{g-f}, \eqref{alpha-zeta}, and
\eqref{appendix-energy-combinations}. In particular, all cross products
generated by the surface radial functions have been retained. The same
substitution gives the compact Coulomb combinations and kinematic functions
below.
The conventional symbols $g_{01}$ and $h_{01}$ are, respectively, the
isotropic and angular electron radial combinations identified explicitly
with $C_{01}$ and $D_{01}$ above; they are not additional independent
functions.
These expressions use the real radial-function convention adopted in
Eq.~\eqref{g-f}; for a convention in which the Coulomb coefficients carry
phases, the products contributing to the decay rate are understood through
their real parts.

For compactness, define
\begin{align}
\alpha_\pm
&\equiv|\alpha_{-1-1}|^2\pm|\alpha_{+1+1}|^2,
&
\beta_\pm
&\equiv|\alpha_{+1-1}|^2\pm|\alpha_{-1+1}|^2,
\nonumber\\
\gamma_+
&\equiv2\operatorname{Re}
(\alpha_{+1+1}\alpha_{-1-1}^*),
&
\delta_+
&\equiv2\operatorname{Re}
(\alpha_{-1+1}\alpha_{+1-1}^*).
\label{coulomb-combinations}
\end{align}
Using Eq.~\eqref{fund-spin} and the radial approximation
\eqref{g-f}, the isotropic kinematic functions in the corrected scalar
$P$-wave convention of Refs.~\cite{PhysRevD.76.093009,PhysRevD.105.099902}
are
\begin{align}
a_{02}^{SP}
&=\frac{\alpha_++\beta_+}{(m_eR)^2},
\nonumber\\
a_{03}^{SP}
&=\frac{1}{3m_eR}
\left[\frac{\zeta}{m_eR}(\alpha_++\beta_+)-2\alpha_-\right],
\nonumber\\
a_{04}^{SP}
&=\frac19\left\{
\left[\left(\frac{\zeta}{2m_eR}\right)^2+1\right]\alpha_+
+\left(\frac{\zeta}{2m_eR}\right)^2\beta_+
-\frac{\zeta}{m_eR}\alpha_-\right\},
\nonumber\\
a_{05}^{SP}
&=\frac{\alpha_++\beta_+}{m_eR},
\nonumber\\
a_{06}^{SP}
&=\frac16\left[
\frac{\zeta}{m_eR}(\alpha_++\beta_+)-2\alpha_-\right]
=\frac{m_eR}{2}a_{03}^{SP}.
\label{a-SP}
\end{align}
The angular-correlation functions are
\begin{align}
b_{02}^{SP}
&=\frac{\gamma_++\delta_+}{(m_eR)^2},
\nonumber\\
b_{03}^{SP}
&=\frac{\zeta}{3m_e^2R^2}(\gamma_++\delta_+),
\nonumber\\
b_{04}^{SP}
&=\frac19\left\{
\left[\left(\frac{\zeta}{2m_eR}\right)^2-1\right]\gamma_+
+\left(\frac{\zeta}{2m_eR}\right)^2\delta_+\right\},
\nonumber\\
b_{05}^{SP}
&=\frac{\gamma_++\delta_+}{m_eR},
\nonumber\\
b_{06}^{SP}
&=\frac{\zeta}{6m_eR}(\gamma_++\delta_+).
\label{a_b-SP}
\end{align}
The normalization of the angular functions follows directly by expanding
$B_0^{SP}$ in Eq.~\eqref{B-SP} with the amplitudes
\eqref{M1-SP}--\eqref{M4-SP}.  In particular, it gives the consistency
relations
\begin{align}
b_{05}^{SP}&=(m_eR)b_{02}^{SP},
&
b_{06}^{SP}&=\frac{m_eR}{2}b_{03}^{SP},
\end{align}
in parallel with
$a_{05}^{SP}=(m_eR)a_{02}^{SP}$ and
$a_{06}^{SP}=(m_eR)a_{03}^{SP}/2$.
Equations~\eqref{a-SP} and \eqref{a_b-SP} generate, respectively, the
isotropic and angular parts of Eq.~\eqref{decay-rate-SP}.  Only the isotropic
functions enter the integrated phase-space factors in
Eq.~\eqref{PSFG_ok-SP}; the angular terms vanish upon integration over the
electron opening angle.

\begingroup
\sloppy

\endgroup

\end{document}